\documentclass[epj]{svjour}
\usepackage{graphics}
\begin{document}
\hugehead
\title{Deeply virtual and exclusive electroproduction of $\omega$ mesons}
%\subtitle{Do you have a subtitle?\\ If so, write it here}
\author{The CLAS collaboration \\
	L. Morand\inst{1} \and 
	D. Dor\'e\inst{1} \and
	M. Gar\c con\inst{1}\thanks{\emph{Corresponding author:} mgarcon@cea.fr} \and
	M. Guidal\inst{2} \and
	J.-M. Laget\inst{1,3} \and
	S. Morrow\inst{1,2} \and
	F. Sabati\'e\inst{1} \and
	E. Smith\inst{3} \and
	G.~Adams\inst{31} \and
P.~Ambrozewicz\inst{11}
% ...
	\\
%	\centerline{(CLAS collaboration)}
%	\centerline{\bf Version after collaboration review, February, 2005}
% \thanks is optional - remove next line if not needed
%\thanks{\emph{Corresponding author:} mgarcon@cea.fr}%
	}                     % End of author list
%
%\offprints{}          % Insert a name or remove this line
%
\institute{
CEA-Saclay, Service de Physique Nucl\'eaire, F91191 Gif-sur-Yvette, France \and 
Institut de Physique Nucl\'eaire, F91406 Orsay, France \and
Thomas Jefferson National Accelerator Facility, Newport News, Virginia 23606, USA \and
Arizona State University, Tempe, Arizona 85287-1504, USA \and 
University of California at Los Angeles, Los Angeles, California 90095-1547, USA \and 
Carnegie Mellon University, Pittsburgh, Pennsylvania 15213, USA \and 
Catholic University of America, Washington, D.C. 20064, USA \and 
Christopher Newport University, Newport News, Virginia 23606, USA \and 
University of Connecticut, Storrs, Connecticut 06269, USA \and 
Edinburgh University, Edinburgh EH9 3JZ, United Kingdom \and 
Florida International University, Miami, Florida 33199, USA \and 
Florida State University, Tallahassee, Florida 32306, USA \and 
Physikalisches Institut der Universit\"at Gie{\ss}en, 35392 Gie{\ss}en, Germany \and 
The George Washington University, Washington, DC 20052, USA \and 
University of Glasgow, Glasgow G12 8QQ, United Kingdom \and 
INFN, Laboratori Nazionali di Frascati, Frascati, Italy \and 
INFN, Sezione di Genova, 16146 Genova, Italy \and 
Idaho State University, Pocatello, Idaho 83209, USA \and 
Institute f\"{u}r Strahlen und Kernphysik, Universit\"{a}t Bonn, Germany \and 
Institute of Theoretical and Experimental Physics, Moscow, 117259, Russia \and 
James Madison University, Harrisonburg, Virginia 22807, USA \and 
Kyungpook National University, Daegu 702-701, The Republic of Korea \and 
Massachusetts Institute of Technology, Cambridge, Massachusetts  02139-4307, USA \and 
University of Massachusetts, Amherst, Massachusetts  01003, USA \and 
Moscow State University, General Nuclear Physics Institute, 119899 Moscow, Russia \and 
University of New Hampshire, Durham, New Hampshire 03824-3568, USA \and 
Norfolk State University, Norfolk, Virginia 23504, USA \and 
Ohio University, Athens, Ohio  45701, USA \and 
Old Dominion University, Norfolk, Virginia 23529, USA \and 
University of Pittsburgh, Pittsburgh, Pennsylvania 15260, USA \and 
Rensselaer Polytechnic Institute, Troy, New York 12180-3590, USA
% ...
}
\date{Received: date / Revised version: date}
% The correct dates will be entered by Springer
%
\abstract{The exclusive $\omega$ electroproduction off the proton was studied
	in a large kinematical domain above the nucleon resonance region
	and for the highest possible photon virtuality ($Q^2$) 
	with the 5.75 GeV beam
	at CEBAF and the CLAS spectrometer. Cross sections were measured 
	up to large values of the four-momentum transfer ($-t<2.7$ GeV$^2$) 
	to the proton.
	The contributions of the interference terms $\sigma_{TT}$ and
	$\sigma_{TL}$ to the cross sections, as well as an analysis of 
%	parameters related to 
	the $\omega$ spin density matrix, indicate that
	helicity is not conserved in this process.
	The $t$-channel $\pi^0$ exchange, or more generally the
	exchange of the associated 
%	saturating 
	Regge trajectory, seems to dominate the
	reaction $\gamma^* p \to \omega p$, even for $Q^2$ as large as 5 GeV$^2$. 
	Contributions of handbag diagrams, related to 
	Generalized Parton Distributions in the nucleon,
	are therefore difficult to extract for this process.
	Remarkably, the high-$t$ behaviour of the cross sections is nearly 
	$Q^2$-independent, which may be interpreted as a coupling of
	the photon to a point-like object in this kinematical limit.
 % end of PACS codes
} %end of abstract
\maketitle
%%%%%%%%%%%%%%%%%%%%%%%%%%%%%%%%%%%%%%%%%%%%%%%%%%%%%%%%%%%%%%%%%%%%
\section{Introduction}
\label{sec:intro}
The exclusive electroproduction of vector mesons is a powerful tool,
on one hand 
to understand the hadronic properties of the virtual photon ($\gamma^*$)
which is exchanged between the electron and the target nucleon~\cite{Bau78},
and on the other hand 
to probe the quark-gluon content of the proton ($p$)~\cite{Str94,Vdh97,Can02}.
At moderate energies in the $\gamma^*p$ system,
but large virtuality of the photon,
quark-exchange mechanisms become significant in the vector meson
production reactions $\gamma^*p\to p\rho$/$\omega$,
thus shedding light on the quark structure of the nucleon.

\begin{figure}
\begin{center}
\resizebox{0.5\textwidth}{!}
	{
	\includegraphics{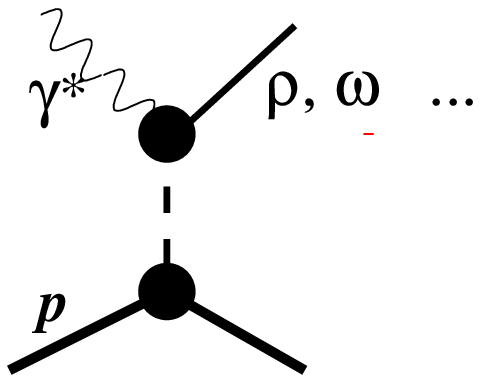} 
	\includegraphics{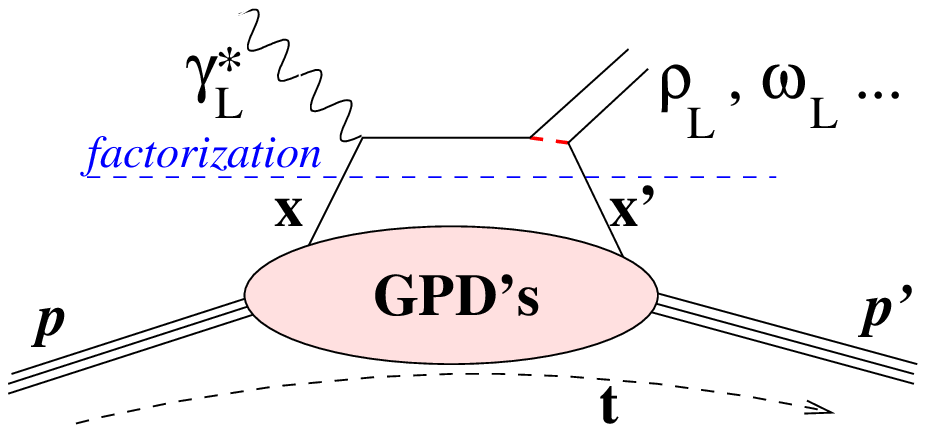}
	}
\caption{Schematic representations of the 
	$t$-channel exchange (left) and of the handbag diagram (right) for 
	exclusive vector meson electroproduction.
	}
\label{fig:handbag} 
\end{center}     
\end{figure}

The interaction of a real photon with nucleons is dominated by its hadronic 
component. The exchange in the $t$-channel of a few Regge trajectories 
permits a description of  
the energy dependence as well as the forward angular 
distribution of many, if not all, real-photon-induced reactions 
(see e.g. Ref.~\cite{Don02}).
For instance, this approach reproduces the photoproduction of vector 
mesons from the CEBAF energy range to the HERA range 
(a few to 200 GeV)~\cite{Lag00}. 
The exchange of the Pomeron (or its realization into two gluons) dominates at high 
energies, while the exchange of meson Regge trajectories ($\pi$, $\sigma$, 
$f_2$) takes over at low energies. At $\gamma p$ energies of a few GeV, 
$\omega$ photoproduction off a proton is dominated by $\pi^0$ exchange 
in the $t$-channel (fig.~\ref{fig:handbag}). 
The use of a saturating Regge trajectory~\cite{Can02} is very 
successful in describing recent photoproduction data~\cite{Bat03}
at large angles 
(large momentum transfer $t$). This is a simple and economical way to 
parameterize hard scattering mechanisms. Extending these measurements to the 
virtual photon sector opens the way to tune the hadronic component of the 
exchanged photon, to explore to what extent $\pi^0$ exchange survives, 
and to observe hard scattering mechanisms with the help of a second hard 
scale, the virtuality $Q^2$ of the photon.

The study of such reactions in the Bjorken   
regime\footnote{$Q^2$ and $\nu$ large and $x_B$ finite, where
	  $-Q^2$ and $\nu$ are the squared mass and the laboratory-frame
	  energy of the virtual photon, while $x_B=Q^2/2M_p\nu$ is the
	  usual Bjorken variable.}
holds promise, through perturbative QCD, to access
the so-called
Generalized Parton Distributions (GPD) of the nucleon~\cite{Ji97,Bel05}.
These structure functions are a generalization of the
parton distributions measured in the deep inelastic scattering
experiments and their first moment links them to the elastic form factors
of the nucleon. Their second moment gives access to the 
sum of the quark spin and the quark orbital angular momentum 
in the nucleon~\cite{Ji97}. 
The process under study may be represented by the so-called handbag
diagram (fig.~\ref{fig:handbag}).
Its amplitude factorizes~\cite{Col97} into a
``hard" process where the virtual photon is absorbed by a quark 
and a ``soft" one containing the new information on the nucleon, the GPD
(which are functions of $x$ and $x'$, the momentum fraction carried
by the quark in the initial and final states, and of $t$, the
squared four-momentum transfer between the initial and final protons).
The factorization applies only to the 
transition, at small values of $-t$, between longitudinal photons ($L$) 
and helicity-0 mesons, which is dominant in the Bjorken regime.
Because of the necessary gluon exchange to produce the
meson in the hard process (see fig.~\ref{fig:handbag}), 
the dominance of the handbag contribution is expected to be reached
at a higher $Q^2$ for meson production than for photon production
(DVCS). Nevertheless, recent results on deeply virtual $\rho$ production
show a qualitative agreement with calculations based on the
handbag diagram~\cite{Air00,Had04}.
Vector meson production
is an important
complement to DVCS, since it singles out the
quark helicity independent GPD $H$ and $E$ which enter
Ji's sum rule~\cite{Ji97} and allows, in principle, 
for a flavor decomposition of these distributions
(see e.g. Ref.~\cite{Die03}).

Apart from early, low statistics, muon production experiments
at SLAC~\cite{Bal74,Pap79}, the leptoproduction
of $\omega$ mesons was measured at DESY~\cite{Joo77},
for $0.3<Q^2<1.4$ GeV$^2$, $W<2.8$ GeV ($x_B<0.3$),
and then at Cornell~\cite{Cas81}, in a wider
kinematical range ($0.7<Q^2<3$ GeV$^2$, $W<3.7$ GeV) but with
larger integration bins. These two experiments yielded cross sections differing
by a factor of about 2 wherever they overlap (around $Q^2\simeq$~1~GeV$^2$).
The DESY experiment also provided the only analysis so far, in electroproduction,
of the $\omega$ spin density
matrix elements, averaged over the whole kinematical range. This analysis
indicated that, in contrast with $\rho$ electroproduction, there
is little increase in the ratio $R$ of longitudinal to transverse cross
sections ($\sigma_L/\sigma_T$) when going from photoproduction to low $Q^2$
electroproduction. More recently, $\omega$
electroproduction was measured at ZEUS~\cite{Bre00},
at high $Q^2$ and very low $x_B$, in a kinematical regime
more sensitive to purely diffractive phenomena and to gluons in the nucleon.
Finally, there is also unpublished data from HERMES~\cite{Tyt01}.

\begin{figure}
\begin{center}
\resizebox{0.45\textwidth}{!}{ \includegraphics{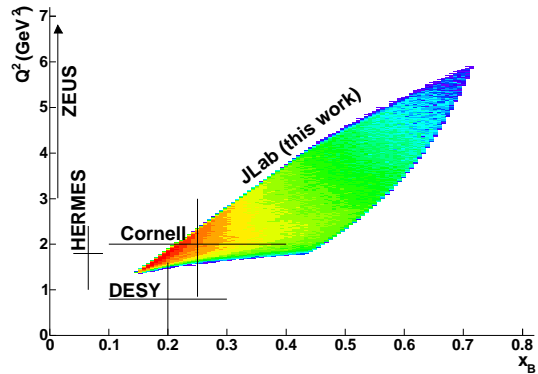}}
\caption{(Color online) Kinematical range covered by this and 
	 previous~\cite{Joo77,Cas81,Bre00,Tyt01} 
	 $\omega$ electroproduction experiments.
	 The lines are indicative of the total coverage in $Q^2$ and $x_B$
	 of previous experiments.
	}
\label{fig:kine} 
\end{center}     
\end{figure}

The main goal of the present
experiment was to reach the highest achievable $Q^2$ values in exclusive
meson electroproduction in the valence quark region.
In the specific case of the $\omega$ production,
it is to test which of the two descriptions --- 
with hadronic or quark degrees of freedom, more specifically 
$t$-channel Regge trajectory exchange or handbag diagram --- 
applies in the considered kinematical domain
(see fig.~\ref{fig:kine}).
For this purpose, the reduced cross sections $\sigma_{\gamma^*p\to \omega p}$
were measured in fine bins in $Q^2$ and $x_B$, as well as their distribution
in $t$ and $\phi$ (defined below). In addition, parameters related 
to the $\omega$ spin density
matrix were extracted from the analysis of the angular
distribution of the $\omega$ decay products.
If the vector meson
is produced with the same helicity as the virtual photon,
$s$-channel helicity conservation (SCHC) is said to hold.
From our results, the relevance of SCHC 
and of natural parity exchange in the $t$-channel
was explored in a model-independent way. 
These properties have been
established empirically in the case of photo- and electroproduction
of the $\rho$ meson (see e.g. Ref.~\cite{Ros71}), 
but may not be a general feature of all
vector meson production channels.

This paper is based on the thesis work of Ref.~\cite{Mor03}, where
additional details on the data analysis may be found.

%%%%%%%%%%%%%%%%%%%%%%%%%%%%%%%%%%%%%%%%%%%%%%%%%%%%%%%%%%%%%%%%%%%%
\section{Experimental procedure}
\label{sec:exp}
\begin{figure}
\begin{center}
\resizebox{0.4\textwidth}{!}{ \includegraphics{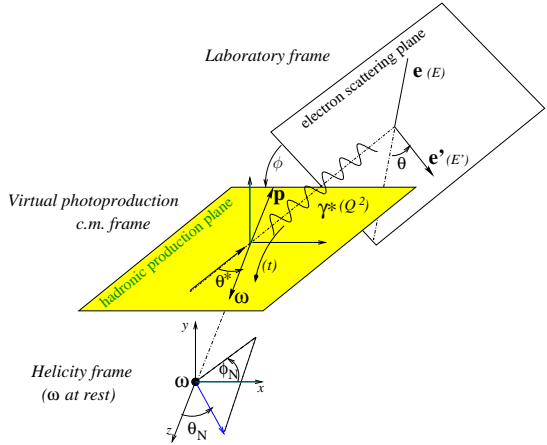}}
\caption{(Color online) Reference frames and relevant variables used for the 
	 description of the reaction $ep\to ep\omega$, followed by
	 $\omega\to\pi^+\pi^-\pi^0$.
	}
\label{fig:ref_frames} 
\end{center}     
\end{figure}

We measured the process $ep\to ep\omega$, followed by the decay 
$\omega\to\pi^+\pi^-\pi^0$. 
The scattered electron and the recoil proton were detected,
together with at least one charged pion from the $\omega$ decay.
At a given beam energy $E$, this process is
described by ten independent kinematical variables. In the absence of
polarization in the $ep$ initial state, the observables are independent of the
electron azimuthal angle in the laboratory.
$Q^2$ and $x_B$ are
chosen to describe the $\gamma^*p$ initial state.
The scattered electron energy $E'$ and,
for ease of comparison with other data,  
the $\gamma^*p$ center-of-mass energy $W$ will be used as well. 
$t$ is the
squared four-momentum transfer from the $\gamma^*$ to the $\omega$, and $\phi$
the angle between the electron ($ee'\gamma^*$) and hadronic
($\gamma^*\omega p$) planes. 
Since $t$ is negative and has a kinematical upper bound $t_0(Q^2,x_B)$
corresponding to $\omega$ production in the direction of the $\gamma^*$,
the variable $t'=t_0-t$ will also be used.
The $\omega$ decay is described in the
so-called helicity frame, where the $\omega$ is at rest and the $z$-axis
is given by the $\omega$ direction in the $\gamma^*p$ center-of-mass
system. In this helicity frame, the normal to the decay plane is characterized by 
the angles $\theta_N$ and $\varphi_N$ (fig.~\ref{fig:ref_frames}).
Finally the distribution of the three pions within the decay plane is
described by two angles and a relative momentum. This latter distribution
is known from the spin and parity of the $\omega$ meson~\cite{Ste62}
and is independent of the $\gamma^*p \to \omega p$ reaction mechanism.
The purpose of the present study is to characterize as completely as possible 
the distributions of cross sections according to the six variables
$Q^2$, $x_B$, $t$, $\phi$, $\cos\theta_N$ and $\varphi_N$.
%So for a given ($Q^2$, $x_B$), and for a given degree of longitudinal
%polarization $\varepsilon(E,E',Q^2)$ of the virtual photon, 
%the event distribution is 
%$\sigma_{\gamma^*p \to \omega p}(\varepsilon,Q^2,x_B,t) \times
%\mathcal{W}(\{r_ij^{\alpha}(\varepsilon,Q^2,x_B,t),\cos\theta_N),\varphi_N,\phi)$

%------------------------------------------------------------------------------------------
\subsection{The experiment}
\label{subsec:expsub}
\begin{figure}
\begin{center}
\resizebox{0.30\textwidth}{!}{ \includegraphics{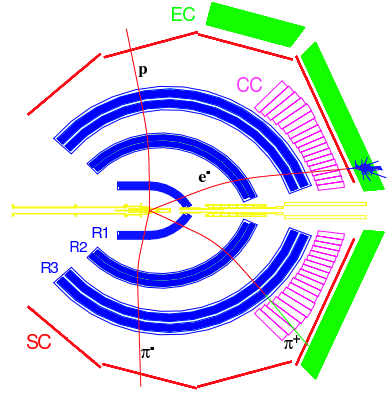}}
\caption{(Color online) Schematic view of the CLAS spectrometer components 
	(see text for description)
	and of typical particle tracks, viewed in projection.
	The torus coils are not shown.
	}
\label{fig:tracksL} 
\end{center}     
\end{figure}

The experiment was performed at the Thomas Jefferson National Accelerator
Facility (JLab).
The CEBAF 5.754 GeV electron beam was directed at a 5-cm long liquid-hydrogen
target. The average beam intensity was 7 nA, resulting in an 
effective integrated luminosity of 28.5 fb$^{-1}$ for the data 
taking period (October 2001 to January 2002). 
The target was positioned at the center of the CLAS spectrometer. 
This spectrometer uses a toroidal magnetic field generated
by six superconducting coils for the determination of particle momenta. 
The field integral varied approximately from 2.2 to 0.5 Tm, 
in average over charges and momenta of different particles,
for scattered angles between 14$^{\circ}$ and 90$^{\circ}$.
All the spectrometer components are arranged in six identical sectors. 
Charged particle trajectories
were detected in three successive packages of drift chambers (DC),
the first one before the region of magnetic field (R1), the second one
inside this region (R2), and the third one after (R3). Threshold \v{C}erenkov 
counters (CC) were used to discriminate pions from electrons.
Scintillators (SC) allowed for a precise determination of the
particle time-of-flight. Finally, a segmented electromagnetic calorimeter (EC)
provided a measure of the electron energy. This geometry and the
event topology are illustrated in fig.~\ref{fig:tracksL}.
A detailed description of the CLAS spectrometer and of its
performance is given in Ref.~\cite{CLAS}.

The data acquisition was triggered by a coincidence CC$\cdot$EC corresponding 
to a minimal scattered electron energy of about 0.575 GeV. The
trigger rate was 1.5 kHz, with a data acquisition dead time of 6\%. 
A total of $1.25\times 10^9$ events was recorded.
%------------------------------------------------------------------------------------------
\subsection{Particle identification}
\label{subsec:pid}
After calibration of all spectrometer subsystems, tracks were reconstructed
from the DC information.
The identification of particles associated with
each track proceeded differently for electrons and hadrons. 

Electrons were identified from the correlation between momentum (from DC)
and energy (from EC). In addition pions were rejected from the electron sample
by a cut in the CC amplitude and imposing a condition on the
energy sharing between EC components compatible with the depth profile
of an electromagnetic shower. Geometrical fiducial cuts ensured that the
track was inside a high efficiency region for both CC and EC. 
The efficiencies of the electron
identification cuts ($\eta_{CC}$ and $\eta_{EC}$) depended on the electron
momentum and angle (or on $Q^2$ and $x_B$). $\eta_{EC}$ was calculated from
data samples using very selective CC cuts in order
to unambiguously select electrons. 
%and looking at the EC detection efficiency for these events. 
$\eta_{CC}$ was extracted from an extrapolation of the 
CC amplitude Poisson distribution into the low amplitude region.
These efficiencies varied respectively between
0.92 and 0.99 (CC), and 0.86 to 0.96 (EC). At low electron energies,
a small contamination of pions remained, which did not however satisfy
the $\omega$ selection criteria to be described below.

The relation between momentum (from DC) and velocity (from path length
in DC and time-of-flight in SC) allowed for a clean identification of 
protons ($p$) and pions ($\pi^+$ and $\pi^-$). However, for momenta
larger than 2 GeV/$c$, ambiguities arose between $p$ and $\pi^+$ identification,
which led us to the discarding of events corresponding to $t<-2.7$ GeV$^2$.
Fiducial cuts were applied to hadrons as well. The efficiency for
the hadron selection cuts was accounted for in the acceptance calculation
described in sect.~\ref{subsec:acc}. 
%------------------------------------------------------------------------------------------
\subsection{Event selection and background subtraction}
\label{subsec:sel}
\begin{figure}
\begin{center}
\resizebox{0.4\textwidth}{!}
	{ \includegraphics{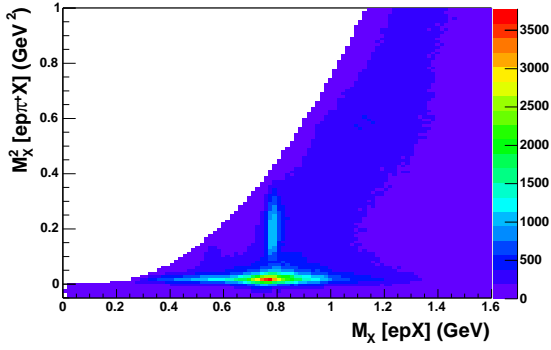}
	}
\caption{(Color online) Identification of the $\omega$ channel in the 
	case of the detection of one charged pion: 
	$M_{X}^2 [ep\pi^+X]$ vs $M_{X} [epX]$.
	Events corresponding to $\rho$ (horizontal locus)
	and other two-pion production channels are clearly
	separated from those corresponding to $\omega$
	production (vertical locus).
	}
\label{fig:mm2dim} 
\end{center}     
\end{figure}

\begin{figure}
\begin{center}
\resizebox{0.5\textwidth}{!}
	{ \includegraphics{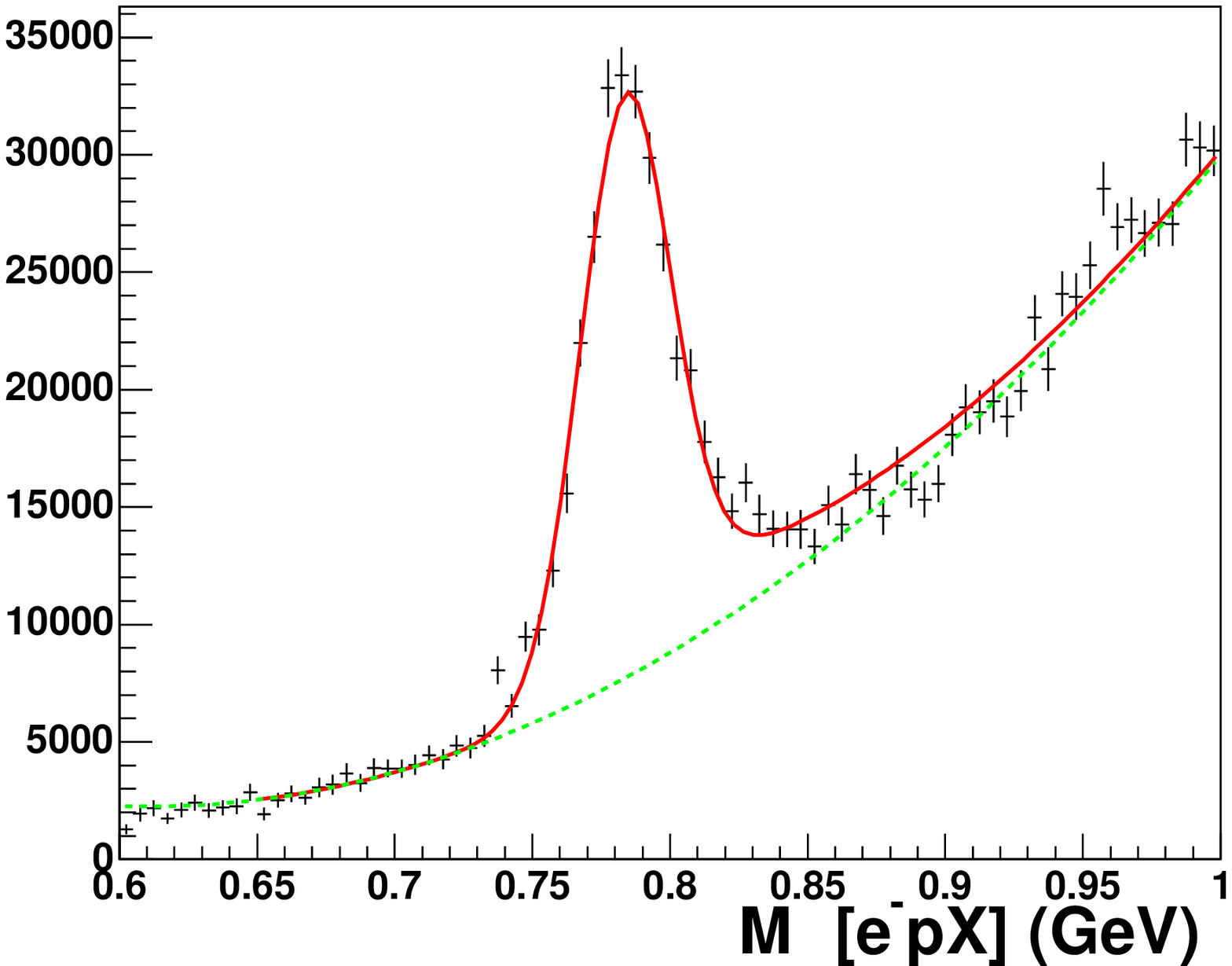}
	  \includegraphics{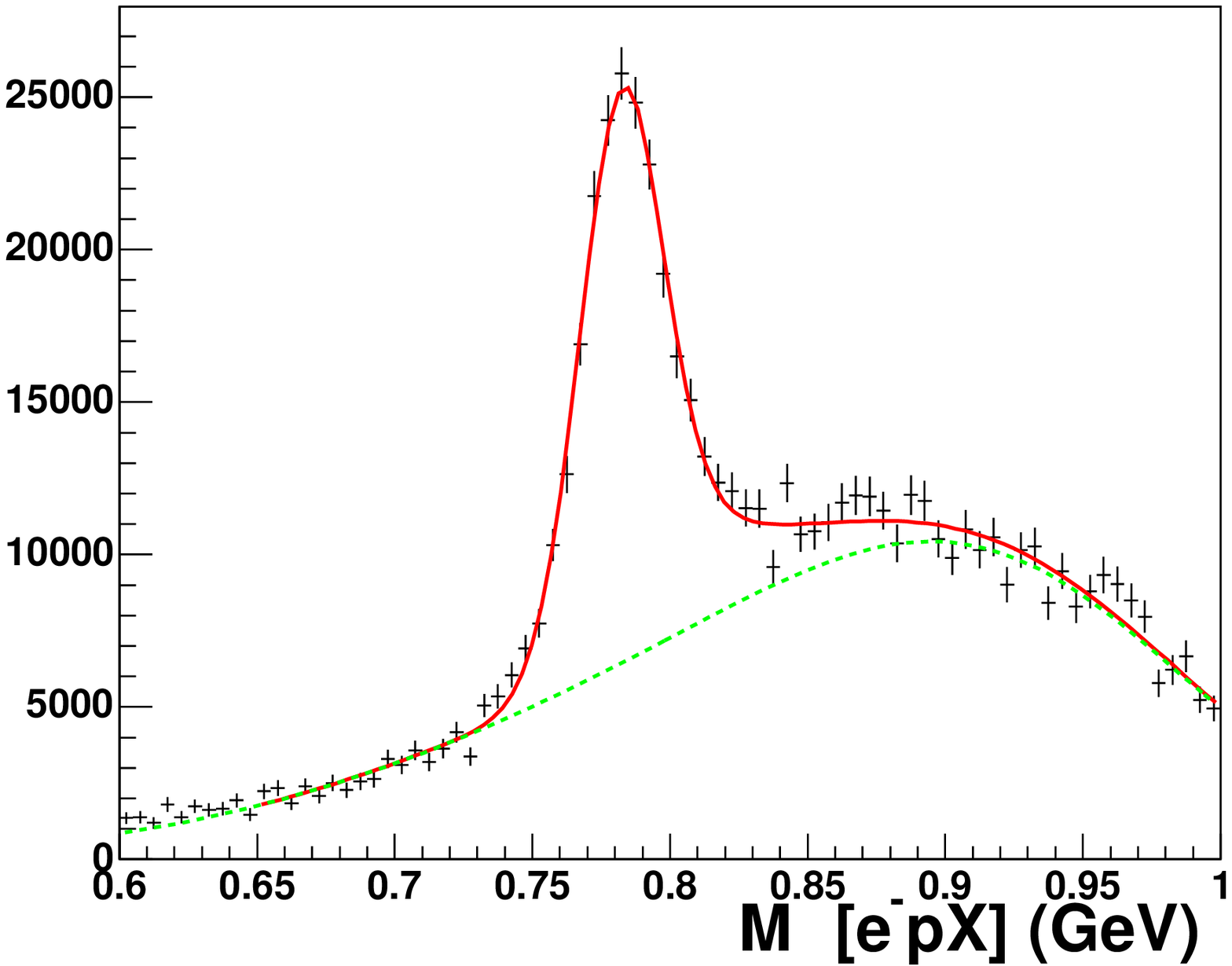}
	}
\caption{(Color online) Missing mass $M_{X} [epX]$ distributions 
	for the $ep\to ep\pi^+X$ event configuration,
	for two ($Q^2,x_B$) bins,
	after selection cuts and event weighting discussed in the text.	
	Left~: 2.2~GeV$^2$~$\leq Q^2 \leq$~2.5~GeV$^2$ and 0.34~$\leq x_B \leq$~0.40. 
	Right~: 3.1~GeV$^2$~$\leq Q^2 \leq$~3.6~GeV$^2$ and 0.52~$\leq x_B \leq$~0.58
	(at the edge of kinematical acceptance). 
	The two lines indicate the subtracted background (green) 
	and the fitted distribution ($\omega$ peak + background in red).
	}
\label{fig:fitexs} 
\end{center}     
\end{figure}

Two configurations of events were studied, with one or two detected charged pions:
$ep\to ep\pi^+X$ and $ep\to ep\pi^+\pi^-X$. 
The former benefits from a larger acceptance and is adequate
to determine cross sections, while the latter is necessary to
measure in addition the distribution of the $\omega$ decay plane
orientation and deduce from it the $\omega$ spin density matrix.
The final
selection of events included cuts in $W$ and $E'$: $W>1.8$ GeV to
eliminate the threshold region sensitive to resonance production~\cite{Bur02} and
$E'>0.8$ GeV to minimize radiative corrections and residual pion
contamination in the electron tracks. The first
configuration was selected requiring a missing mass $M_X$ larger than 0.316 GeV
to eliminate two pion production channels 
($M_X^2>0.1$ GeV$^2$ on the vertical axis of fig.~\ref{fig:mm2dim}). 
This cut was chosen slightly
above the two pion mass in order to minimize background. The corresponding
losses in $ep\to ep\omega$ events were very small and accounted for
in the acceptance calculation to be discussed below.
Events corresponding to the $\omega$ production appear as a clear peak in
the $ep\to epX$ missing mass spectrum (fig.~\ref{fig:fitexs}). The width
of this peak ($\sigma\simeq 16$ MeV) is mostly due to the experimental resolution.

After proper weighting of each event with the acceptance calculated as
indicated in sect.~\ref{subsec:acc},  
a background subtraction was performed
for each of 34 bins ($Q^2,x_B$) and, for
differential cross sections, for each bin in $t$ or $\phi$. 
The background was determined by a fit to the acceptance-weighted 
distributions with a second-order polynomial and a peak shape as
modeled by simulations (a skewed gaussian shape taking into account the
experimental resolution and radiative tail). At the smallest values of
$W$, the fitted background
shape was modified to account for kinematical acceptance cuts. 
The acceptance-weighted numbers of $ep \to ep\omega$ events 
% ($n_{w}(Q^2,x_B)$ in eq.~(\ref{eq:sigma}))
were computed using the sum of
weighted counts in the $M_X[epX]$ distributions for $.72<M_X<.85$ GeV,
diminished by the fitted background integral in the same interval.

\begin{figure}
\begin{center}
\resizebox{0.4\textwidth}{!}
	{ \includegraphics{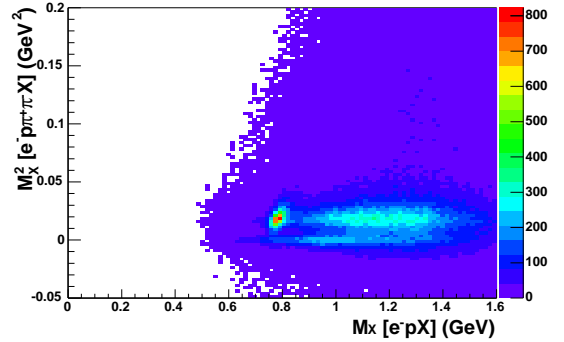}
	}
\caption{(Color online) Identification of the $\omega$ channel in the 
	case of the detection of two charged pions: 
	$M_{X}^2 [ep\pi^+\pi^-X]$ vs $M_{X} [epX]$ 
	for $M_{X}^2 [ep\pi^{\pm}X] \geq$~0.1~GeV$^2$. 
%	$M_{X} [ep\pi^-X] \geq$~0.1~GeV$^2$. 
	The spot at (~0.78, 0.02) corresponds to the $\omega$.
	}
\label{fig:mm2dim2} 
\end{center}     
\end{figure}

\begin{figure}
\begin{center}
\resizebox{0.4\textwidth}{!}
	{ \includegraphics{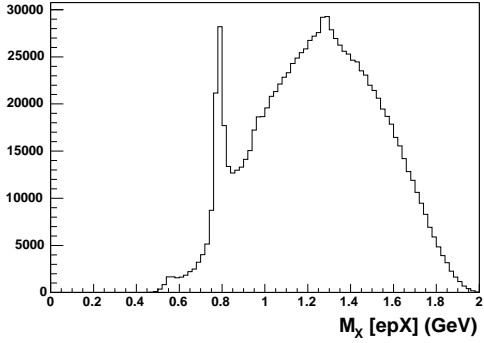}
	}
\caption{Unweighted $M_X[epX]$ spectrum for all $ep\to ep\pi^+\pi^-X$ events
	after selection cuts discussed in the text.
	}
\label{fig:mmep} 
\end{center}     
\end{figure}

Likewise, events from the second configuration  ($ep\to ep\pi^+\pi^-X$) were
selected with cuts in missing masses: \\ 
$M_X[ep\pi^+X]$ and $M_X[ep\pi^-X]>0.316$ GeV, 
$0.01 \ \hbox{GeV}^2 \leq M_{X}^2 [ep\pi^+\pi^-X] \leq 0.045 \ \hbox{GeV}^2$
(fig.~\ref{fig:mm2dim2}). The resulting \\
$M_X[epX]$ spectrum
after these cuts is illustrated in fig.~\ref{fig:mmep}. 
The background subtraction in the spectrum of weighted events
proceeded in the same way  for each of  64
($Q^2$, $x_B$, $\cos\theta_N$) bins 
or for each of 64 ($Q^2$, $x_B$, $\varphi_N$) bins
in order to analyze the $\omega$ decay distribution
(see sect.~\ref{sec:decay}). 


%------------------------------------------------------------------------------------------
\subsection{Acceptance calculation}
\label{subsec:acc}
The tracks reconstruction
and the event selection were simulated using
a GEANT-based Monte Carlo (MC) simulation of the CLAS spectrometer. 
%, including radiative effects as discussed in sect.~\ref{subsec:rad}. 
We used an
event generator tuned to reproduce photoproduction and low $Q^2$ data
in the resonance region and extrapolated into our kinematical domain~\cite{genev}.
The acceptance was defined in each elementary bin in all relevant variables as
the ratio of accepted  to
generated MC events. At the limit of small six-dimensional bins, 
it is independent of the model used to generate the MC events.
The MC simulation included a tuning
of the DC and SC time resolutions to reproduce the observed widths of the
hadron particle identification spectra and of the missing mass spectra,
so that the efficiency of the corresponding cuts described above 
could be correctly determined.

For the extraction of cross sections from the  $ep\to ep\pi^+X$
configuration, acceptance calculations were performed in 1837
four-dimensional bins ($Q^2$, $x_B$, $t$ and $\phi$) with two
different assumptions about the event distribution
in $\cos\theta_N$ and $\varphi_N$.
The two different MC calculations were used for
an estimate of the corresponding systematic uncertainties 
(see sect.~\ref{subsec:xsec_syst}).
For the analysis of the decay plane distribution
$\mathcal{W}(\cos\theta_N,\varphi_N,\phi)$ 
from the $ep\to ep\pi^+\pi^-X$ configuration, the
acceptance calculation was performed in 3575 six-dimensional bins ($Q^2$,
$x_B$, $t$, $\phi$, $\cos\theta_N$ and $\varphi_N$).
The binning is defined in table~\ref{tab:bin} 
and the numbers above correspond to kinematically allowed bins
that have significant statistics.

\begin{figure}
\begin{center}
\resizebox{0.5\textwidth}{!}
	{ \includegraphics{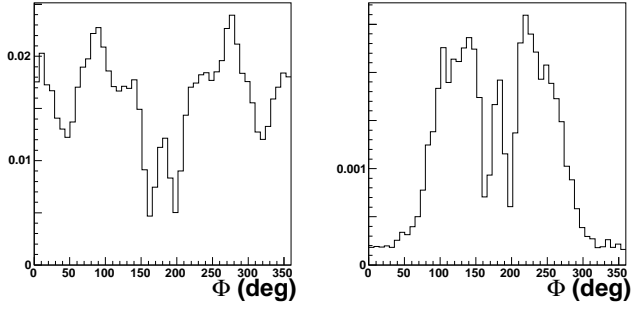}
	}
%	\vspace{3.5cm}
\caption{$\phi$ dependence of calculated acceptance, integrated over other kinematical
	variables, for the one and two detected pion configurations:
	$ep \to ep\pi^+X$ (left), $ep \to ep\pi^+\pi^-X$ (right).
	}
\label{fig:phi_acc} 
\end{center}     
\end{figure}

The calculated acceptances are, on average, of the order of 2\% and 0.2\% respectively for
the two event configurations of interest. They vary smoothly for all variables
except $\phi$, where oscillations, due to the dead zones in the CLAS sectors, 
reproduce the physical distributions of events (fig.~\ref{fig:phi_acc}).
Each event was then weighted with the inverse of the corresponding acceptance. 
Events belonging to bins with either very large or poorly determined  weights 
were discarded
(for the  $ep\to ep\pi^+X$ configuration, acceptance smaller than 0.25\% or associated MC 
statistical uncertainty larger than 35\%). The corresponding losses (a few percent) were 
quantified through the MC efficiency $\eta_{MC}$ by applying these cuts to MC events
and computing the ratio of weighted accepted MC events to generated events.
No attempt was made to calculate the acceptance for the non-resonant
three-pion background, so that the background
shape in fig.~\ref{fig:fitexs} differs from the physical distribution
$d\sigma/dM_X$ when $M_X$ differs from the $\omega$ mass.

%------------------------------------------------------------------------------------------
\subsection{Radiative corrections}
\label{subsec:rad}
Radiative corrections were calculated following Ref.~\cite{Mo69}. They were
dealt with in two separate steps. The MC acceptance calculation presented above
took into account radiation losses due to the emission of hard photons,
through the application of the cut $M_X[epX]<0.85$ GeV. Corrections
due to soft photons, and especially the virtual processes arising from
vacuum polarization and vertex correction, were determined separately for
each bin in ($Q^2$, $x_B$, $\phi$). The same event generator employed for the
computation of the acceptance was used, with radiative effects turned on and
off, thus defining a corrective factor $F_{rad}$. The $t$-dependence of $F_{rad}$
is smaller than all uncertainties discussed in sect.~\ref{subsec:xsec_syst} 
and was neglected.

%%%%%%%%%%%%%%%%%%%%%%%%%%%%%%%%%%%%%%%%%%%%%%%%%%%%%%%%%%%%%%%%%%%%
\section{Cross sections for $\gamma^*p\to\omega p$}
\label{sec:xsec}
The total reduced cross sections 
% for a given state of the $\gamma^*p$ system
were extracted from the data through~:
\begin{eqnarray}
\sigma_{\gamma^* p \rightarrow \omega p} (Q^2, x_B, E) &=&
\frac{1}{\Gamma_V(Q^2,x_B,E)} \times 
\frac{n_{w}(Q^2,x_B)}{B\mathcal{L}_{int} \Delta Q^2 \cdot \Delta x_B} \nonumber \\
& &\times \frac{F_{rad}}{\eta_{CC} \ \eta_{EC} \ \eta_{MC}}\ .
\label{eq:sigma}
\end{eqnarray}

The Hand convention~\cite{Han63} was used for the definition of the virtual 
transverse photon flux $\Gamma_V$, which includes here
a Jacobian in order to express the cross sections in the chosen
kinematical variables~:
\begin{equation}
\Gamma_V (Q^2,x_B,E) = \frac{\alpha}{8\pi} \frac{Q^2}{M_p^2 E^2} 
\frac{1-x_B}{x_B^3} \frac{1}{1-\varepsilon},
\label{eq:GammaV}
\end{equation}
with the virtual photon polarization parameter being defined as~:
\begin{equation}
\varepsilon = \frac{1}{1+2\frac{Q^2+(E-E')^2}{4EE'-Q^2}}\ .
\label{eq:epsilon}
\end{equation}

In eq.~(\ref{eq:sigma}), $n_{w}(Q^2,x_B)$ is the acceptance-weighted number of
$ep\to ep\omega$ events after background subtraction.
The branching ratio of the $\omega$ decay into three pions is $B=0.891$~\cite{PDG04}.
The integrated effective lumimosity $\mathcal{L}_{int}$ includes the data acquisition
dead time correction. 
$\Delta Q^2$ and $\Delta x_B$ are the corresponding bin widths;
for bins not completely filled (because of $W$ or $E'$ cuts on the electron,
or of detection acceptance), the phase space $\Delta Q^2\cdot\Delta x_B$
includes a surface correction and the $Q^2$ and $x_B$ central
values are modified accordingly. The radiative correction factor and the
various efficiencies not included in the MC calculation were discussed
in previous sections.

Differential cross sections in $t$ or $\phi$ were extracted in a similar manner.
Cross section data and corresponding MC data for the acceptance calculation were
binned according to table~\ref{tab:bin}. 
\begin{table}
\begin{center}
\caption{Definition of binning for cross section (1) 
         and $\omega$ polarization (2) data.
         N refers to the number of bins in the
         specified range for each variable.
         }
\label{tab:bin}       
\begin{tabular}{lcccc}
\hline\noalign{\smallskip}
Variable        & Range(1)   &N(1)& Range(2)   & N(2) \\
\noalign{\smallskip}\hline\noalign{\smallskip}
$Q^2$ (GeV$^2$) & 1.6 - 3.1  & 5  & 1.7 - 4.1  & 4 \\
                & 3.1 - 5.1  & 4  & 4.1 - 5.2  & 1 \\
$x_B$           & 0.16 - 0.64 & 8  & 0.18 - 0.62& 4 \\
$-t$ (GeV$^2$)  & 0.1 - 1.9  & 6  & 0.1 - 2.1  & 4 \\
                & 1.9 - 2.7  & 1  & 2.1 - 2.7  & 1 \\
$\phi$ (rd) & 0 - 2$\pi$ & 9 & 0 - 2$\pi$      & 6/9/12 \\
$\cos\theta_N$  & -          & -  & $-1$ - 1     & 8 \\
$\varphi_N$ (rd) & -         & -  & 0 - 2$\pi$ & 8 \\    
\noalign{\smallskip}\hline
\end{tabular}
\end{center}
\end{table}
%------------------------------------------------------------------------------------------
\subsection{Systematic uncertainties}
\label{subsec:xsec_syst}
Systematic uncertainties in the cross section measurements arise from the determination
of the CLAS acceptance, of electron detection efficiencies, of the luminosity,
and from the background subtraction. They are listed in table~\ref{tab:systerr1}
and discussed hereafter.

Errors in the acceptance calculation may
be due to inhomogeneity in the detectors response, such as faulty channels in
DC or SC, to possible deviations between experimental and simulated resolutions
in spectra where cuts were applied, to the input of the event generator
(both in cross section and in decay distribution $\mathcal{W}_{gen}$), to radiative
corrections and finally to the event weighting procedure. The most significant
of these uncertainties (8\%) was quantified by performing 
a separate complete MC simulation
varying inputs for the parameters describing the decay distribution 
$\mathcal{W}_{gen}$.

Systematic uncertainties on the electron detection efficiencies were estimated with
experimental data, by varying the electron selection cuts or the extrapolated
CC amplitude distribution (see sect.~\ref{subsec:pid}).

Systematic background subtraction uncertainties were estimated by varying the assumed
background functional shapes. In particular, the background curvature under the
$\omega$ peak was varied between extreme values compatible 
with an equally good
fit to the distributions in fig.~\ref{fig:fitexs}. For bins corresponding to
low values of $W$, the acceptance cut to the right of the $M_X[epX]$ peak
induced an additional uncertainty.

Finally, overall normalization uncertainties were due to the knowledge of target
thickness (2\%) and density (1\%) and of beam integrated charge (2\%).

Errors
contributing point-to-point and to the overall normalization are separately
added in quadrature in table~\ref{tab:systerr1}.
For $t$ or $\phi$ distributions, the same uncertainties apply,
but may contribute to the overall normalization uncertainty instead of point-to-point. 
For example, the
shape of the $M_X[epX]$ distributions depends mostly on $Q^2$ and $x_B$, not
on $t$ and $\phi$; the background subtraction uncertainties are then considered
as a normalization uncertainty for the $d\sigma/dt$ and $d\sigma/d\phi$
distributions. The uncertainties on $\eta_{MC}$ are largest for $t$ and $\phi$ bins
with the smallest acceptance (lowest and highest $t$ values, as well as
$\phi\simeq 180^{\circ}$).

\begin{table}	
\begin{center}
\caption{Point-to-point and normalization systematic uncertainties, for integrated
	and differential cross sections.
	} 	
\begin{tabular}{llll}
\hline\noalign{\smallskip}
Source of uncertainty                              & $\sigma$            & $d\sigma/dt$ & $d\sigma/d\phi$ \\
\noalign{\smallskip}\hline\noalign{\smallskip}
{\it CLAS acceptance}                        &                     &              &     \\
- inhomogeneities                            & 6\%                 & 6\%          & 6\% \\
- resolutions                                & 2\%                 & 2\%          & -   \\
- $\sigma_{gen}(Q^2,x_B,t)$                  & 5\%                 & 5\%          & -   \\
- $\mathcal{W}_{gen}(\cos\theta_N,\varphi_N,\phi)$ & 8\%                 & 8\%    & -   \\
- radiative corrections                      & 4\%                 & -            & 2\% \\
- binning                                    & 5\%                 & 5\%          & -   \\
- $\eta_{MC}$                                 & 4\%                 & 2-7\%        & 2-20\% \\
{\it Electron detection}                     &                     &              &     \\
- $\eta_{CC}$                                 & 1.5\%               & -            & -   \\
- $\eta_{EC}$                                 & 2\%                 & -            & -   \\		
{\it Background subtraction}                 & 7-11\%              & -            & -   \\
\noalign{\smallskip}\hline\noalign{\smallskip}	
Point-to-point                         & 16-18\%             & 13-14\%      & 7-21\%      \\
\noalign{\smallskip}\hline\noalign{\smallskip}
Normalization 			     & 3\%                 & 9-12\%    & 14-16\%     \\
\noalign{\smallskip}\hline	
\end{tabular}	
\label{tab:systerr1}
\end{center}	
\end{table}	

%------------------------------------------------------------------------------------------
\subsection{Integrated reduced cross sections}
\label{subsec:xsec_int}
\begin{figure}
\begin{center}
\resizebox{0.46\textwidth}{!}
	{ \includegraphics{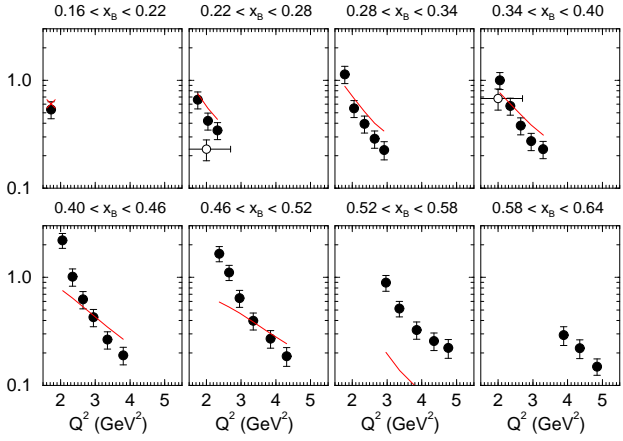}
	}
\caption{(Color online) Reduced cross sections $\gamma^* p \rightarrow \omega p$ 
	as a function of $Q^2$ for differents bins in $x_B$, 
	in units of $\mu$b. Full circles: this work; open circles:
	Ref.~\cite{Cas81}. The red cross and curves correspond to the 
	JML model~\cite{Lag04} discussed in sect.~\ref{sec:regge}.
	}
\label{fig:xsec_Q2xB} 
\end{center}     
\end{figure}

\begin{figure}
\begin{center}
\resizebox{0.4\textwidth}{!}
	{ \includegraphics{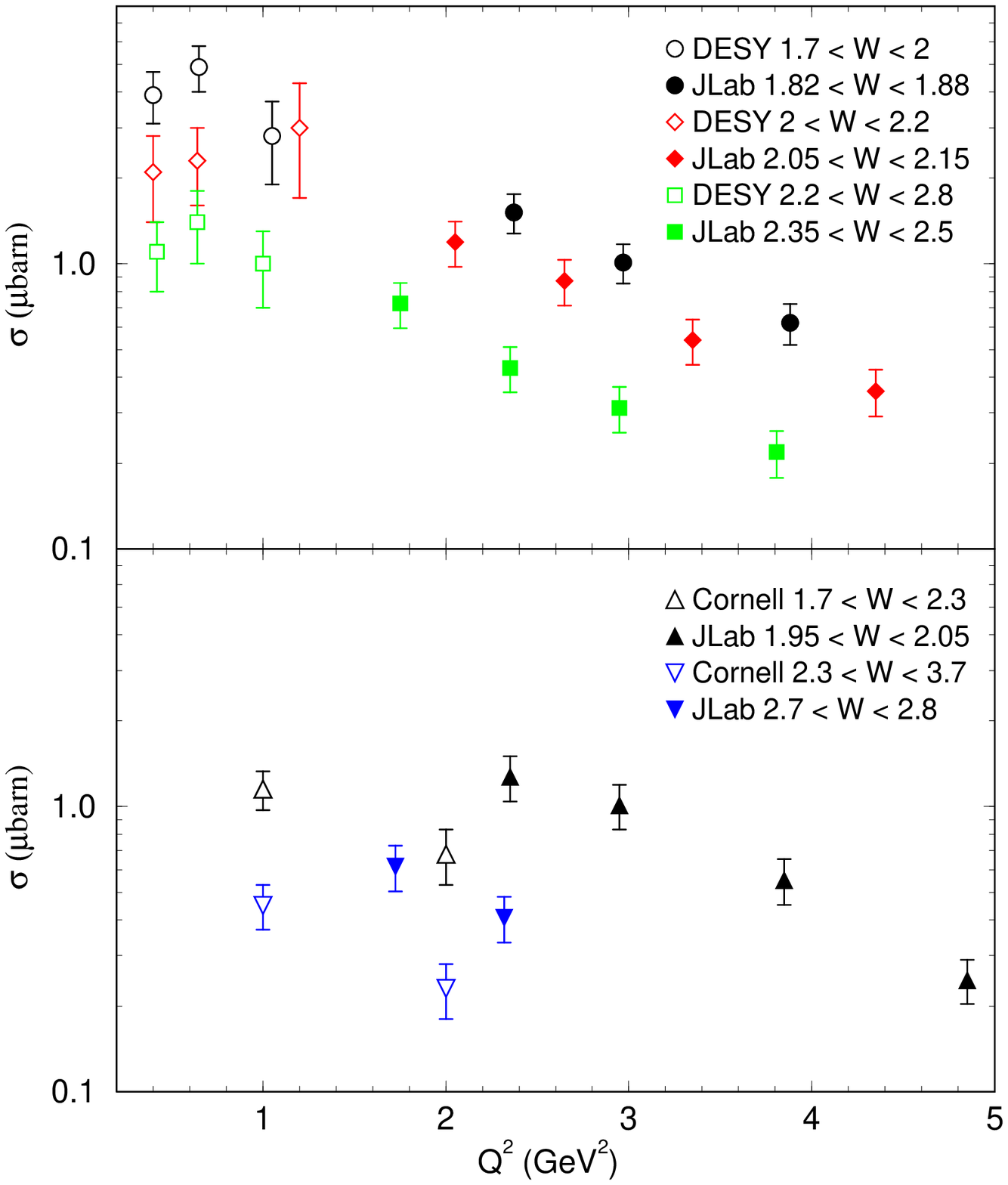}
	}
\caption{(Color online) 
	Total cross sections 
	for the reaction $\gamma^* p \to \omega p$,
	as a function of $Q^2$ and at fixed $W$~: 
	this work in full symbols, 
	DESY~\cite{Joo77} (top) and Cornell~\cite{Cas81} (bottom) in open symbols.
	Each symbol (or color) corresponds to a given central value of $W$~(GeV).
	Note the range of integration in $W$ for each data set.
%	{\it Will be modified.}
	}
\label{fig:sigmaWfixe} 
\end{center}     
\end{figure}

Results for $\sigma_{\gamma^*p\to\omega p}(Q^2,x_B)$ are given in 
table~\ref{tab:sigma} and 
fig.~\ref{fig:xsec_Q2xB}.
For the purpose of comparison with previous data,
fig.~\ref{fig:sigmaWfixe} shows cross sections as a function of $Q^2$ for 
fixed, approximately constant, values of $W$.
When comparing different data sets, note that 
$\sigma = \sigma_T + \varepsilon\sigma_L$ depends on the 
beam energy through $\varepsilon$. However, as will be shown, it is likely that
the difference of longitudinal contributions between two different
beam energies $(\varepsilon_2-\varepsilon_1)\sigma_L$ is much smaller than
the total cross section $\sigma$. 
In addition, the range of integration in $t$ is different for all experiments,
larger in this work, 
but most of the total cross section comes from small $-t$ values.
A direct comparison of the cross sections is then meaningful.

\begin{table*}	
\begin{center}
\caption{Cross sections $\sigma = \sigma_T + \varepsilon\sigma_L$ 
	 and interference terms $\sigma_{TT}$ and $\sigma_{TL}$
	 for the reaction $\gamma^* p \to \omega p$,
	 integrated over $-2.7\ \hbox{GeV}^{2}<t<t_0$.   
	 Slope $b$ of $d\sigma/dt$ for 
	 $-1.5\ \hbox{GeV}^{2} <t<t_0$. 
	 Quoted uncertainties are obtained from the addition in quadrature 
	 of statistical uncertainties
	 and of point-to-point systematic uncertainties as discussed in 
	 sect.~\ref{subsec:xsec_syst}.
	 }
\begin{tabular}{ccccccccc}
\hline\noalign{\smallskip}	
$x_{B}$ & $Q^2$ & $W$ & $\varepsilon$ & $t_0$ 
	& $\sigma\pm\Delta\sigma$& $\sigma_{TT}\pm\Delta\sigma_{TT}$ 
        & $\sigma_{TL}\pm\Delta\sigma_{TL}$ 
        & $b\pm db$ \\
&(GeV$^2$)& (GeV) & & (GeV$^2$) & (nb) &           (nb)  &          (nb)  &        (GeV$^{-2}$)   \\                                   	
\noalign{\smallskip}\hline\noalign{\smallskip}	
0.203 & 1.725 &  2.77 &  0.37 & -0.09 &  536 $\pm$  96&   60 $\pm$  87&    2 $\pm$  30&  2.44 $\pm$ 0.18 \\
 0.250 & 1.752 &  2.48 &  0.59 & -0.15 &  661 $\pm$ 118&  156 $\pm$  61&  -35 $\pm$  24&  1.93 $\pm$ 0.16 \\
 0.252 & 2.042 &  2.63 &  0.43 & -0.14 &  421 $\pm$  75&  104 $\pm$  47&  -18 $\pm$  18&  2.28 $\pm$ 0.16 \\
 0.265 & 2.320 &  2.70 &  0.32 & -0.14 &  344 $\pm$  62&   58 $\pm$  74&  -14 $\pm$  26&  1.88 $\pm$ 0.17 \\
 0.308 & 1.785 &  2.21 &  0.72 & -0.25 & 1139 $\pm$ 205&  310 $\pm$ 122& -175 $\pm$  60&  1.23 $\pm$ 0.17 \\
 0.310 & 2.050 &  2.33 &  0.63 & -0.23 &  551 $\pm$  98&  121 $\pm$  42&  -66 $\pm$  20&  1.90 $\pm$ 0.16 \\
 0.310 & 2.350 &  2.47 &  0.50 & -0.21 &  395 $\pm$  71&  103 $\pm$  41&  -49 $\pm$  17&  2.03 $\pm$ 0.16 \\
 0.313 & 2.639 &  2.58 &  0.37 & -0.20 &  287 $\pm$  52&  111 $\pm$  48&   -9 $\pm$  18&  1.90 $\pm$ 0.17 \\
 0.327 & 2.914 &  2.62 &  0.28 & -0.22 &  226 $\pm$  43&  138 $\pm$  84&  -46 $\pm$  27&  1.87 $\pm$ 0.23 \\
 0.370 & 2.050 &  2.09 &  0.74 & -0.37 & 1002 $\pm$ 180&   91 $\pm$  75&  -25 $\pm$  39&  0.97 $\pm$ 0.17 \\
 0.370 & 2.350 &  2.21 &  0.65 & -0.34 &  581 $\pm$ 104&  150 $\pm$  48&  -36 $\pm$  23&  1.35 $\pm$ 0.17 \\
 0.370 & 2.650 &  2.32 &  0.55 & -0.31 &  380 $\pm$  68&   85 $\pm$  40&  -33 $\pm$  17&  1.62 $\pm$ 0.17 \\
 0.370 & 2.950 &  2.43 &  0.43 & -0.30 &  273 $\pm$  49&   95 $\pm$  42&  -44 $\pm$  17&  1.93 $\pm$ 0.18 \\
 0.378 & 3.295 &  2.51 &  0.31 & -0.30 &  230 $\pm$  42&   17 $\pm$  55&  -27 $\pm$  19&  1.46 $\pm$ 0.18 \\
 0.429 & 2.055 &  1.90 &  0.81 & -0.59 & 2203 $\pm$ 348&   54 $\pm$ 158&  143 $\pm$  85&  1.03 $\pm$ 0.25 \\
 0.430 & 2.350 &  2.00 &  0.74 & -0.53 & 1013 $\pm$ 182&  181 $\pm$  82& -102 $\pm$  43&  0.78 $\pm$ 0.22 \\
 0.430 & 2.650 &  2.10 &  0.67 & -0.49 &  626 $\pm$ 113&   90 $\pm$  63&  -41 $\pm$  32&  0.81 $\pm$ 0.22 \\
 0.430 & 2.950 &  2.19 &  0.58 & -0.46 &  427 $\pm$  78&   56 $\pm$  48&  -24 $\pm$  23&  0.92 $\pm$ 0.23 \\
 0.430 & 3.350 &  2.31 &  0.45 & -0.43 &  265 $\pm$  48&  105 $\pm$  37&  -12 $\pm$  15&  1.34 $\pm$ 0.23 \\
 0.436 & 3.807 &  2.41 &  0.30 & -0.42 &  191 $\pm$  35&  128 $\pm$  60&  -54 $\pm$  21&  1.47 $\pm$ 0.24 \\
 0.481 & 2.371 &  1.85 &  0.79 & -0.79 & 1660 $\pm$ 265&  -34 $\pm$ 173&  241 $\pm$  97&  0.92 $\pm$ 0.50 \\
 0.490 & 2.651 &  1.91 &  0.74 & -0.75 & 1113 $\pm$ 177&  291 $\pm$ 109&   15 $\pm$  62&  0.68 $\pm$ 0.37 \\
 0.490 & 2.950 &  1.99 &  0.68 & -0.69 &  644 $\pm$ 116&  174 $\pm$  69& -154 $\pm$  35&  0.23 $\pm$ 0.34 \\
 0.490 & 3.350 &  2.09 &  0.58 & -0.64 &  397 $\pm$  72&   84 $\pm$  49&  -49 $\pm$  24&  1.39 $\pm$ 0.25 \\
 0.490 & 3.850 &  2.21 &  0.43 & -0.60 &  272 $\pm$  50&   72 $\pm$  46&    5 $\pm$  20&  1.41 $\pm$ 0.26 \\
 0.494 & 4.307 &  2.30 &  0.29 & -0.58 &  187 $\pm$  37&  169 $\pm$  79&  -14 $\pm$  27&  0.78 $\pm$ 0.32 \\
 0.538 & 2.968 &  1.85 &  0.73 & -0.99 &  894 $\pm$ 148&  111 $\pm$ 148&    8 $\pm$  69&  $-$ \\
 0.549 & 3.357 &  1.91 &  0.66 & -0.96 &  514 $\pm$  84&   83 $\pm$  67&  -12 $\pm$  31&  $-$ \\
 0.550 & 3.850 &  2.01 &  0.55 & -0.88 &  327 $\pm$  59&   95 $\pm$  54&  -52 $\pm$  26&  0.65 $\pm$ 0.49 \\
 0.550 & 4.350 &  2.11 &  0.41 & -0.83 &  258 $\pm$  48&   29 $\pm$  62&  -13 $\pm$  27&  0.90 $\pm$ 0.43 \\
 0.557 & 4.765 &  2.16 &  0.31 & -0.83 &  222 $\pm$  44&  -42 $\pm$ 121&   55 $\pm$  51&  1.36 $\pm$ 0.67 \\
 0.601 & 3.882 &  1.86 &  0.61 & -1.26 &  292 $\pm$  57&   91 $\pm$  89&   74 $\pm$  50&  $-$ \\
 0.610 & 4.352 &  1.91 &  0.52 & -1.24 &  221 $\pm$  43&   75 $\pm$  54&  -53 $\pm$  23&  $-$ \\
 0.610 & 4.850 &  2.00 &  0.40 & -1.16 &  150 $\pm$  26& -110 $\pm$  48&   39 $\pm$  20&  $-$ \\
\noalign{\smallskip}\hline	
\end{tabular}	
\label{tab:sigma} 
\end{center}	 
\end{table*} 

There is no direct overlap between the present data and the DESY data~\cite{Joo77}, 
but they seem to be compatible with a common trend. The Cornell 
data~\cite{Cas81} are roughly
a factor 2 lower than ours. Where they overlap, the Cornell data are also
a factor 2 lower than the DESY data. We can only make the following
conjectures as to the origin of this discrepancy: 
the Cornell results do not appear to have been
corrected for internal virtual radiative effects (about 15\%); their overall 
systematic uncertainty in absolute cross sections is 25\%; their acceptance calculation
has a model dependence which was not quantified and in particular the 
decay distribution, given in eq.~(\ref{eq:omedec}) below, was assumed flat; 
finally, the estimate of average values of kinematical variables 
$\langle Q^2 \rangle$ and $\langle W \rangle$ may be an additional source 
of uncertainty since the 
corresponding bins are at least 5 times larger than in the present work.

%------------------------------------------------------------------------------------------
\subsection{$t$ dependence of cross sections}
\label{subsec:xsec_t}
\begin{figure}
\begin{center}
\resizebox{0.35\textwidth}{!}
	{ \includegraphics{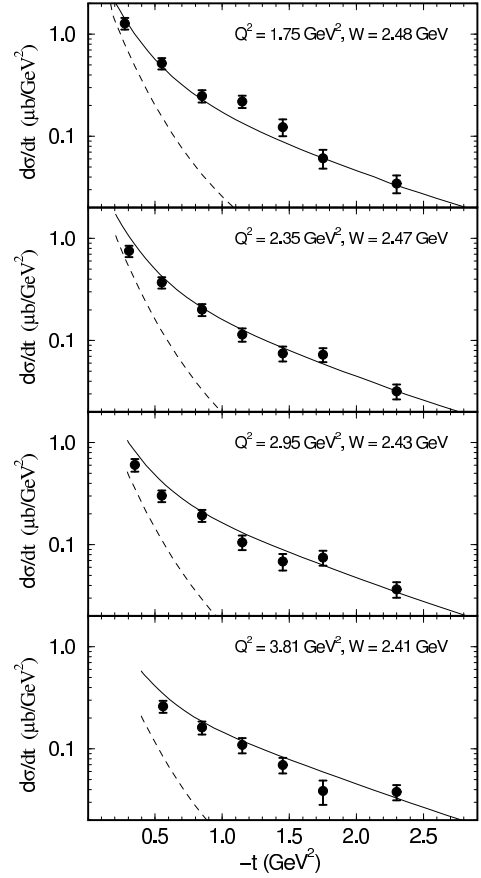}
	}
\caption{$d\sigma/dt$ 
	for the reaction $\gamma^* p \to \omega p$,
	at $W$~$\simeq$~2.45~GeV for different bins in $Q^2$~: 
	our data and the JML model (discussed in sect.~\ref{sec:regge})
	with  $F_{\pi\omega\gamma}$ given by
	eq.~(\ref{eq:Fpiomegg2}) (full lines) and without the
	$t$-dependence in this equation
	(dashed lines). 
%	{\it Should we combine the four curves on the same graph ??}
	}
\label{fig:dsigmadtWfixe} 
\end{center}
\end{figure}

\begin{figure}
\begin{center}
\resizebox{0.35\textwidth}{!}
	{ \includegraphics{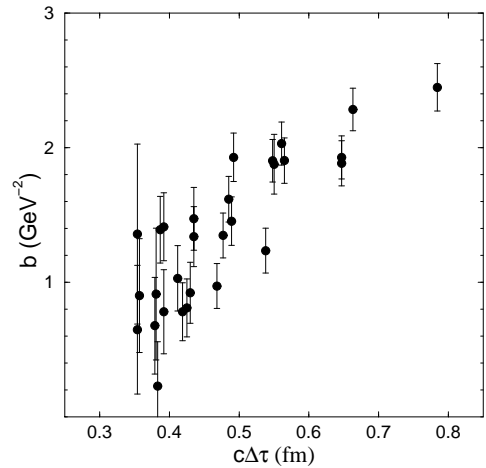}
	}
\caption{Slope $b$ of $d\sigma/dt$,
	for the reaction $\gamma^* p \to \omega p$,
%	determined over $-1.5\ \hbox{ GeV}^2<t<t_0$, 
	as a function of the
	formation length $c\Delta\tau$, cf. eq.~(\ref{eq:cdt}).
%	{\it Will be improved.}
	}
\label{fig:pente} 
\end{center}
\end{figure}

\begin{figure}
\begin{center}
\resizebox{0.35\textwidth}{!}
	{ \includegraphics{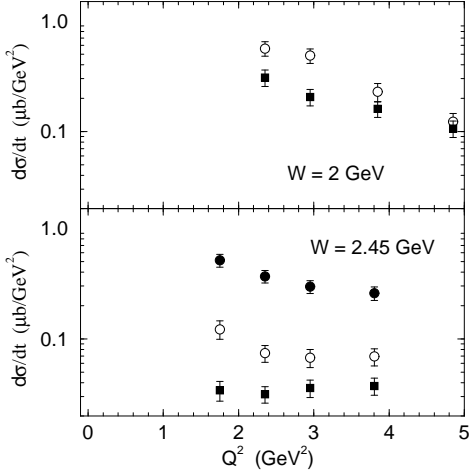}
	}
\caption{$d\sigma/dt$ at fixed values of $t$ and $W$, as a function of $Q^2$,
	for the reaction $\gamma^* p \to \omega p$~:
	$t=-0.55$ (full circles), $t=-1.45$ (empty circles) and
	$t=-2.30$ (squares) GeV$^2$.
%	{\it Add photoproduction data?}
	}
\label{fig:dsigmadt_vs_Q2} 
\end{center}
\end{figure}

Four of the 34 distributions of differential cross sections
are illustrated in fig.~\ref{fig:dsigmadtWfixe}. 
The general features of these distributions are of a diffractive type
($d\sigma/dt \propto e^{bt}$) at small values of $-t$. 
The values of the slope $b$, as determined from a fit of the
distributions in the interval -1.5 GeV$^2<t<t_0$, are
between 0.5 and 2.5 GeV$^{-2}$ and are compiled in
table~\ref{tab:sigma}.
They are also plotted as a function of the formation length
(distance of fluctuation of the virtual photon in a real meson)~\cite{Bau78}:
\begin{equation}
c\Delta\tau =\frac{1}{\sqrt{\nu^2+Q^2+M_{\omega}^2}-\nu}
\label{eq:cdt}
\end{equation}
in fig.~\ref{fig:pente}.
They are compatible with those obtained for the reaction
$\gamma^* p \rightarrow \rho^0 p$~\cite{Had04}.
There exists only one previous determination~\cite{Cas81} of this
quantity for $\gamma^* p \to \omega p$, 
integrated over a wide kinematical range
corresponding to $0.6<c\Delta\tau<2.5$ fm, with a value 
$b= 6.1\pm 0.8$ GeV$^{-2}$. 
This larger value of $b$ is consistent with the observed
discrepancy between the Cornell experiment and the present work.
For larger values of $-t$, 
the slope of the cross sections becomes much smaller 
and $d\sigma/dt$ becomes nearly independent of $Q^2$,
except for the lowest values of $W$ (see fig.~\ref{fig:dsigmadt_vs_Q2}).
This is certainly a new finding from this experiment,
which may indicate
a point-like coupling of the virtual photon 
to the target constituents in this kinematical regime.
This behaviour will be discussed quantitatively in 
Sect.~\ref{sec:regge}.
%------------------------------------------------------------------------------------------
\subsection{$\phi$ dependence of cross sections}
\label{subsec:xsec_phi}
\begin{figure}
\begin{center}
\resizebox{0.5\textwidth}{!}
	{ \includegraphics{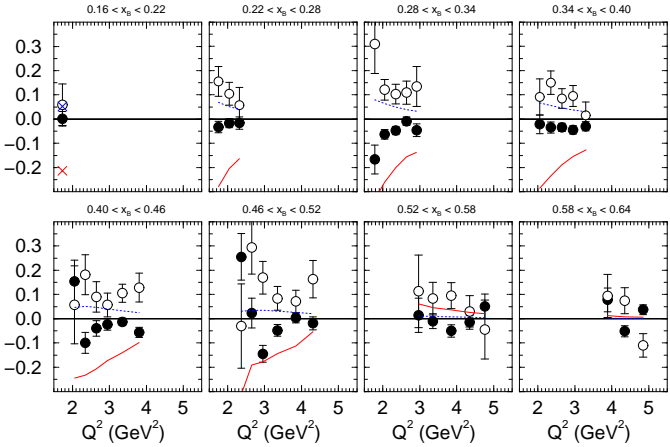}
	}
\caption{(Color online) $\sigma_{TT}$ (open circles) and $\sigma_{TL}$ 
	(full circles), in units of $\mu$b, 
	for the reaction 
	$\gamma^* p \rightarrow \omega p$ as a function of $Q^2$ for different 
	bins in $x_B$, integrated over $-2.7\ \hbox{GeV}^{2}<t<t_0$.
	The dashed blue and full red curves are the corresponding
	calculations in the JML model~\cite{Lag04} 
	discussed in sect.~\ref{sec:regge}.
	}
\label{fig:sigma_tt_tl} 
\end{center}
\end{figure}

The 34 $\phi$ distributions have the expected  $\phi$ dependence~:
\begin{equation}
\frac{d\sigma}{d\phi} = \frac{1}{2\pi} \left( \sigma + \varepsilon \cos 2 \phi \ \sigma_{TT} + \sqrt{2\varepsilon(1+\varepsilon)} \cos \phi \ \sigma_{TL} \right).
\label{eq:sigmaphi}
\end{equation}
The interference terms $\sigma_{TT}$ and $\sigma_{TL}$ were extracted
from a fit of each distribution with eq.~(\ref{eq:sigmaphi}).
The results appear in fig.~\ref{fig:sigma_tt_tl} and in 
table~\ref{tab:sigma}.
If helicity were conserved in the $s$-channel (SCHC), these
interference terms $\sigma_{TT}$ and $\sigma_{TL}$ would vanish.
It does not appear to be the case in fig.~\ref{fig:sigma_tt_tl}.
The $\phi$ distributions do not support the SCHC hypothesis.

%%%%%%%%%%%%%%%%%%%%%%%%%%%%%%%%%%%%%%%%%%%%%%%%%%%%%%%%%%%%%%%%%%%%
\section{Analysis of $\omega$ decay distribution}
\label{sec:decay}
In the absence of polarization in the initial state, 
the distribution of the pions from $\omega$ decay is characterized by 
eq.~(\ref{eq:omedec})~\cite{Sch73}.
%\begin{floatequation}
%\mbox{\textit{see equation~\ref{eq:omedec} above}}
%\end{floatequation}
\begin{figure*}
%\begin{eqnumber}{3}
\begin{eqnarray}
\mathcal{W}(\cos \theta_N,\varphi_N, \phi) &=& \frac{3}{4\pi} \displaystyle\left[ \frac{1}{2}(1-r_{00}^{04}) + \frac{1}{2}(3r_{00}^{04}-1)\cos^2 \theta_N
-\sqrt{2} \mbox{Re} r_{10}^{04} \sin 2\theta_N \cos \varphi_N - r_{1-1}^{04} \sin^2 \theta_N \cos 2\varphi_N  \right. \nonumber \\
&& \hspace*{7mm} -\varepsilon \cos 2\phi ( r_{11}^1 \sin^2 \theta_N + r_{00}^1 \cos^2 \theta_N - \sqrt{2} \mbox{Re} r_{10}^1 \sin 2\theta_N \cos \varphi_N - r_{1-1}^1 \sin^2 \theta_N \cos 2\varphi_N) \nonumber \\
&& \hspace*{7mm} -\varepsilon \sin 2\phi ( \sqrt{2} \mbox{Im} r_{10}^2 \sin 2\theta_N \sin \varphi_N + \mbox{Im} r_{1-1}^2 \sin^2 \theta_N \sin 2\varphi_N) \nonumber \\
&& \hspace*{7mm} + \sqrt{2\varepsilon(1+\varepsilon)} \cos \phi ( r_{11}^5 \sin^2 \theta_N + r_{00}^5 \cos^2 \theta_N 
 - \sqrt{2} \mbox{Re} r_{10}^5 \sin 2\theta_N \cos \varphi_N -r_{1-1}^5 \sin^2 \theta_N \cos 2\varphi_N) \nonumber \\
&& \hspace*{7mm} \left. + \sqrt{2\varepsilon(1+\varepsilon)} \sin \phi ( \sqrt{2} \mbox{Im} r_{10}^6 \sin 2\theta_N \sin \varphi_N + \mbox{Im} r_{1-1}^6 \sin^2 \theta_N \sin 2\varphi_N) 
\displaystyle\right] \label{eq:omedec}\\
&& \hspace*{-35mm} 
\mbox{where the parameters } r_{ij}^{\alpha}\mbox{, hereafter referred to as 
	matrix elements, are related to the }
\omega \mbox{ spin density matrix:} \nonumber \\
r_{ij}^{04} &=& \frac{\rho_{ij}^0 + \varepsilon R \rho_{ij}^4}{1+\varepsilon R} ; 
\hspace*{1cm}
r_{ij}^{\alpha} = \frac{\rho_{ij}^{\alpha}}{1+\varepsilon R}\mbox{ for } \alpha = 1,2\ ; 
\hspace*{1cm}
r_{ij}^{\alpha} = \sqrt{R} \frac{\rho_{ij}^{\alpha}}{1+\varepsilon R}\mbox{ for }\alpha = 5,6 . 
\label{eq:rhotor}
\end{eqnarray}
\smallskip
\rule{\textwidth}{.4pt}
\end{figure*} 
%where $\varepsilon$ is the virtual photon polarization parameter and 
%$R$ the ratio $\sigma_L/\sigma_T$. 
The quantities $\rho_{ij}^{\alpha}$ are defined from a decomposition of the
$\omega$ spin density matrix on a basis of 9
hermitian matrices. The superscript 
$\alpha$ refers to this decomposition and it is related to the virtual photon 
polarization ($\alpha=0$--2 for transverse photons, 
$\alpha=4$ for longitudinal photons, 
and $\alpha=5$--6 for interference between $L$ and $T$ terms). 
For example, $\rho_{00}^0$ is related to the probability of 
the transition between a transverse photon and a longitudinal meson.

All elements 
$\rho_{ij}^{\alpha}$ can be expressed as bilinear combinations of 
helicity amplitudes 
which describe the $\gamma^*p \to \omega p$ transition~\cite{Don02,Sch73}. 
An analysis of the $\mathcal{W}$ distribution
can then be used to test whether helicity is conserved
in the $s$-channel (SCHC), that is between the virtual photon and 
$\omega$. If SCHC applies, $\rho_{00}^0=0$ 
and $\rho_{00}^4=1$. Then eq.~(\ref{eq:rhotor}) leads to a direct relation
between the measured $r_{00}^{04}$ and the ratio $R=\sigma_L/\sigma_T$. In that case,
the longitudinal and transverse cross sections
may be extracted from data without a Rosenbluth separation.

\begin{figure}
\begin{center}
\resizebox{0.47\textwidth}{!}
	{ \includegraphics{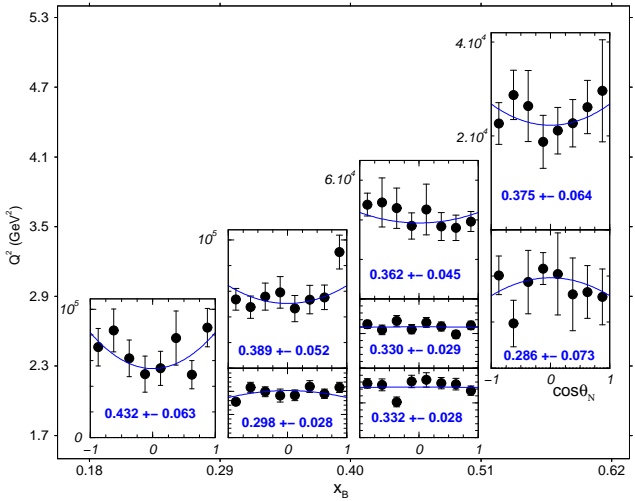}
	}
\caption{(Color online) Distributions of acceptance-weighted
	and background-subtracted counts as a function of $\cos\theta_N$, 
	for 8 bins in ($Q^2$, $x_B$).
	The location and size of each graph correspond to the
	($Q^2$, $x_B$) range over which the data is integrated.
	On all graphs, one division on the vertical axis
	represents $2\times 10^4$ (arbitrary units).
	All data are integrated in $t$ ($-t<2.7$ GeV$^2$).  
	The blue curves correspond to fits with eq.~(\ref{eq:WcosthetaN}),
	with the resulting 
 	$r_{00}^{04}$ and its statistical uncertainty indicated on each distribution.
 	The systematic uncertainty on this matrix element is estimated at 0.042. 
 	}
\label{fig:fitsCOSthetaN}
\end{center}
\end{figure}

\begin{figure}
\begin{center}
\resizebox{0.47\textwidth}{!}
	{ 
	\includegraphics{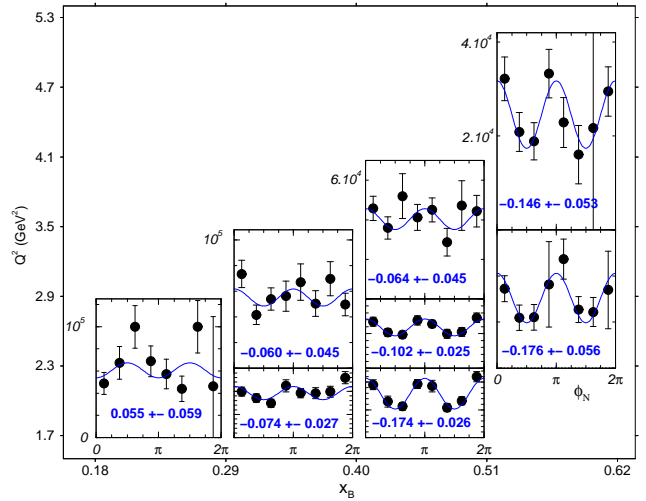}
	}
\caption{(Color online) Distributions of acceptance-weighted
	and background-subtracted counts as a function of $\varphi_N$.
	The blue curves correspond to fits with eq.~(\ref{eq:WvarphiN}),
	with the resulting 
 	$r_{1-1}^{04}$ and its statistical uncertainty indicated on each distribution.
 	The systematic uncertainty on this matrix element is estimated at 0.042.
	See also legend of fig.~\ref{fig:fitsCOSthetaN}.
	} 
\label{fig:fitsVARphiN}
\end{center}
\end{figure}

The matrix elements $r_{00}^{04}$ and $r_{1-1}^{04}$ were first extracted
using one-dimensional projections of the $\mathcal{W}$ distribution.
Note that $r_{1-1}^{04}$ should be zero if SCHC applies.
Integrating eq.~(\ref{eq:omedec}) over $\phi$ and then respectively
over $\varphi_N$ or $\cos\theta_N$, one gets:
\begin{eqnarray}
\mathcal{W}(\cos\theta_N) &=& \frac{3}{4} 
    \left[(1-r_{00}^{04})+(3r_{00}^{04}-1)\cos^2\theta_N \right],
         \label{eq:WcosthetaN} \\
\mathcal{W}(\varphi_N) &=& \frac{1}{2\pi} \left[1-2r_{1-1}^{04}\cos 2\varphi_N \right].
         \label{eq:WvarphiN}
\end{eqnarray}
The background subtraction in the $\cos\theta_N$ or $\varphi_N$ distributions 
was performed in 8 bins of the corresponding variables, and for 8 
bins in ($Q^2$, $x_B$). The
number of acceptance-weighted events was extracted from the corresponding 
$M_{X}[epX]$ distribution, as in sect.~\ref{subsec:sel}. See
figs.~\ref{fig:fitsCOSthetaN} and~\ref{fig:fitsVARphiN} for results,
together with fits to eqs.~(\ref{eq:WcosthetaN}) and~(\ref{eq:WvarphiN})
respectively.

\begin{figure}
\begin{center}
\resizebox{0.47\textwidth}{!}
	{ \includegraphics{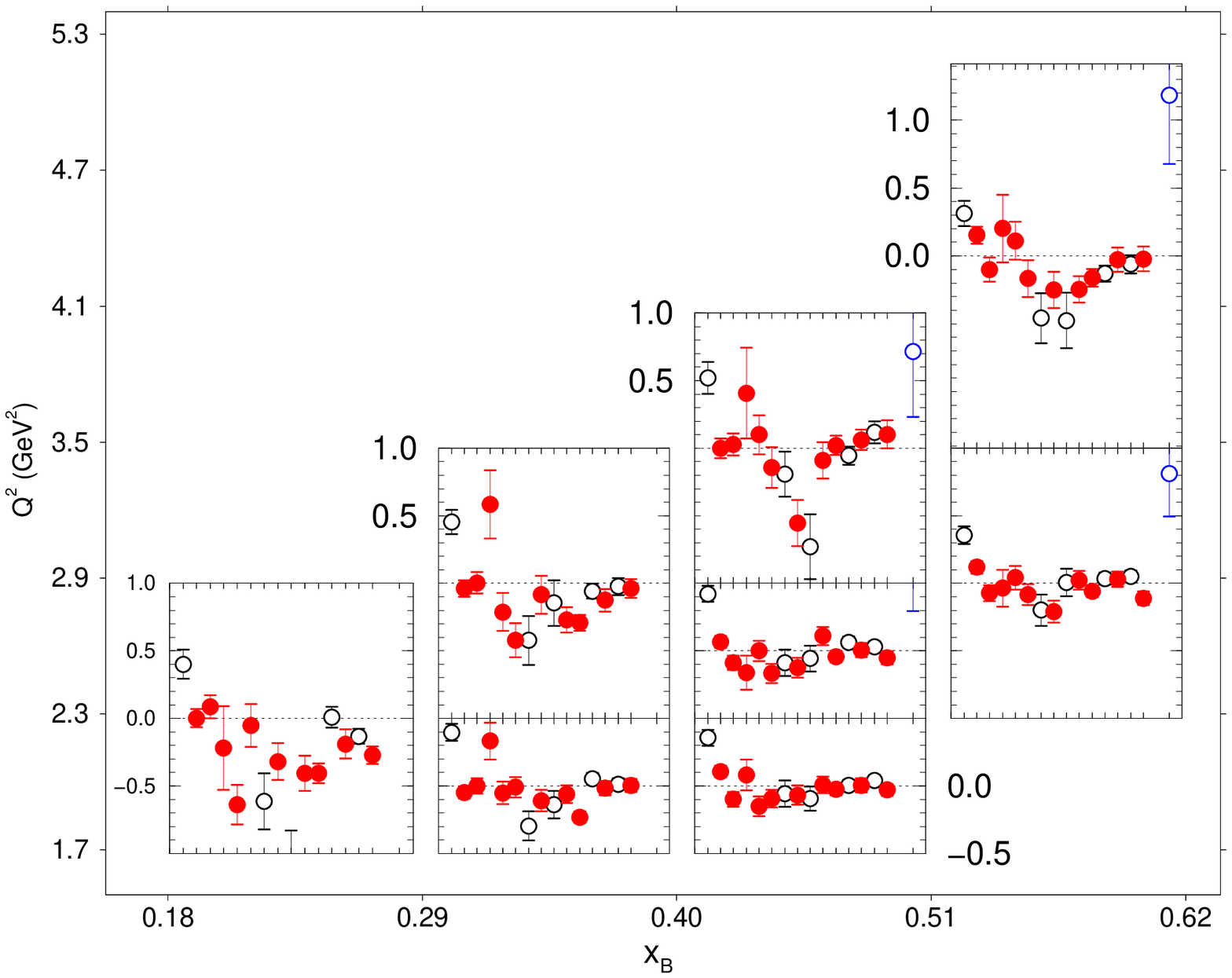}
	}
\caption{(Color online) $r_{ij}^{\alpha}$ extracted with the method of moments for
	8 bins in ($Q^2$, $x_B$) and for $t'<0.5$ GeV$^2$.
	The location and size of each graph correspond to the
	($Q^2$, $x_B$) range over which the data is integrated, but the
	scale is the same on all graphs. 
	The abscissa on each graph corresponds to the following list of matrix elements:
	$r_{00}^{04}$, 
	Re~$r_{10}^{04}$, $r_{1-1}^{04}$, $r_{00}^{1}$, $r_{11}^{1}$, Re~$r_{10}^{1}$, 
	$r_{1-1}^{1}$, Im~$r_{10}^2$, Im~$r_{1-1}^2$, $r_{00}^{5}$, $r_{11}^{5}$, 
	Re~$r_{10}^{5}$, $r_{1-1}^{5}$, Im~$r_{10}^{6}$, Im~$r_{1-1}^{6}$.
	The red filled symbols indicate those matrix elements which are zero if SCHC
	applies. The 16th entry (blue empty circle, in some cases off scale) 
	is the combination of $r_{ij}^{\alpha}$ given by eq.~(\ref{eq:NPE}). 
	Error bars include systematic uncertainties added in quadrature.
	} 
\label{fig:rijMOMmeth3}
\end{center}
\end{figure}

Alternatively, the 15 matrix elements $r_{ij}^{\alpha}$ may be expressed 
in terms of moments of the decay distribution \\ 
$\mathcal{W}(\cos\theta_N,\varphi_N,\phi)$~\cite{Sch73}.
This method of expressing moments includes the background contribution under
the $\omega$ peak (about 25\%). It yields compatible results with
the (background subtracted) 1D projection method for $r_{00}^{04}$ and $r_{1-1}^{04}$.
It was used to study the $t$ dependence of
$r_{ij}^{\alpha}$  and to evaluate the systematic uncertainties in their
determination. 
Results for forward $\gamma^*p\to\omega p$ reaction ($t'<0.5$ GeV$^2$)
are given in fig.~\ref{fig:rijMOMmeth3}.
Systematic uncertainties originate from the determination of the
MC acceptance. The main source of uncertainties was found to be the 
finite bin size in $\phi$. Calculations with different bin sizes
(see table~\ref{tab:bin}) and checks of higher, unphysical,
moments in the event distribution led to systematic uncertainties of 0.02
to 0.08, depending on the $r_{ij}^{\alpha}$ matrix element. In addition,
cuts in the event weights were varied,
resulting in a systematic uncertainty of about 0.03 for all matrix elements.

Finally, the $r_{ij}^{\alpha}$ matrix elements were also extracted
using an unbinned maximum likelihood method. Results were compatible
with the first two methods. In view of the
$\phi$ dependence of the acceptance
(see fig.~\ref{fig:phi_acc}), this method was used for checking the
validity of the $r_{ij}^{\alpha}$ determination when restricting the
$\phi$ range taken into consideration in the fit.

These studies lead to the conclusion that SCHC does not hold for the
reaction $\gamma^*p\to\omega p$, not only when considering the
whole $t$ range (fig.~\ref{fig:fitsVARphiN}), but also, though
in a lesser extent, in the forward direction (fig.~\ref{fig:rijMOMmeth3}).
For SCHC, all matrix elements become zero, except five: 
$r_{00}^{04}$, $r_{1-1}^{1}$, Im~$r_{1-1}^2$, Re~$r_{10}^{5}$, Im~$r_{10}^{6}$
and these are not all independent; 
they satisfy~\cite{Don02}: 
$r_{1-1}^{1}=-$Im$r_{1-1}^2$ and Re$r_{10}^{5}=-$Im$r_{10}^{6}$.
The quantity
\begin{eqnarray}
\chi^2 = \frac{1}{12}
	\left[
	\sum_1^{10} \left(\frac{r}{\Delta r}\right)^2
	\right.
	&&+ \frac{(r_{1-1}^{1}+\mbox{Im} r_{1-1}^2)^2}{(\Delta r_{1-1}^{1})^2+(\Delta\mbox{Im} r_{1-1}^2)^2}  
%	\right. 
	\nonumber \\
	+&& 
	\left.
	\frac{(\mbox{Re}r_{10}^{5}+\mbox{Im} r_{10}^6)^2}{(\Delta \mbox{Re}r_{10}^{5})^2+(\Delta\mbox{Im} r_{10}^6)^2}
	\right]
\end{eqnarray}
where the sum is carried over the ten matrix elements which would be zero if
SCHC applies, may be used as a measure of SCHC violation. Including
in the denominators $\Delta r$ the systematic uncertainties added in quadrature
to the statistical uncertainties, the 7 $\chi^2$ values 
(excluding the distributions at the lowest $x_B$ bin where 
SCHC violation is the most manifest in fig.~\ref{fig:rijMOMmeth3})
range from 2.3 to 7.7 when including all data, and
drop only to 1.7 to 5.1, in spite of doubled statistical uncertainties,
when considering only the forward production ($t'<0.5$ GeV$^2$).
Furthermore, when examining the relation between these matrix elements
and helicity-flip amplitudes, it does not appear possible to ascribe
the SCHC violation to a small subset of these amplitudes.
It is therefore not justified to calculate $R$ from eq.~(\ref{eq:rhotor})
and separate the longitudinal and transverse cross sections from this data.

When one retains only those amplitudes which correspond to a natural
parity exchange in the $t$-channel, then the following relation should
hold~\cite{Tyt01}~:
\begin{equation}
1 - r_{00}^{04} + 2 r_{1-1}^{04} - 2 r_{11}^{1} - 2 r_{1-1}^{1} = 0
\label{eq:NPE}
\end{equation}
This particular combination is plotted as the 16th point on each of
the graphs of fig.~\ref{fig:rijMOMmeth3}. 
The fact that it is not zero points to the
importance of the unnatural parity (presumably pion) exchange.

It is also possible to estimate qualitatively the role of pion exchange
through the U/N asymmetry of the transverse cross section, where U and N
refer to unnatural and natural parity exchange contributions~\cite{Don02}:
\begin{equation}
P \equiv \frac{\sigma_T^N - \sigma_T^U}{\sigma_T^N + \sigma_T^U} 
			= (1+\varepsilon R)(2r_{1-1}^1 - r_{00}^1)\ .
\label{eq:PUN}
\end{equation}
Our results yield $r_{1-1}^1<0$ and $r_{00}^1\geq 0$ over the whole 
kinematical range, and thus:
%Since $|\rho_{ij}^1|>|r_{ij}^1|$ (see eq.~(\ref{eq:rhotor})), we are led to~:
\begin{equation}
P < - ( 2|r_{1-1}^1| + |r_{00}^1| )\ .
\end{equation}
Hence $P$ is large and negative, which means that most of the transverse cross
section is due to unnatural parity exchange.

%%%%%%%%%%%%%%%%%%%%%%%%%%%%%%%%%%%%%%%%%%%%%%%%%%%%%%%%%%%%%%%%%%%%
\section{Comparison with a Regge model}
\label{sec:regge}

Regge phenomenology was applied with success to the photoproduction of
vector mesons in our energy range and at higher energies~\cite{Can02,Sib03}.
Laget and co-workers showed that the introduction of saturating Regge
trajectories provides an excellent simultaneous description of 
the high $-t$ behaviour of the $\gamma p \to p \rho,\omega,\phi$ cross sections,
given an appropriate choice of the relevant coupling constants.
The $t$-channel exchanges considered in this JML model are
indicated in table~\ref{tab:JML}. 
Saturating trajectories have a close 
phenomenological connection to the quark-antiquark interaction which governs
the mesonic structure~\cite{Ser94}. They provide an effective way
to implement gluon exchange between the quarks forming the exchanged meson.
\begin{table}
\begin{center}
\caption{Meson and Pomeron (or two-gluon) exchanges considered in the JML
	 model for vector meson production.}
\label{tab:JML}       
\begin{tabular}{cc}
\hline\noalign{\smallskip}
Produced         & Exchanged  \\
vector meson        & Regge trajectories \\
\noalign{\smallskip}\hline\noalign{\smallskip}
$\rho$ & $\sigma$, $f_2$, P/2g \\
$\omega$           &  $\pi^0$, $f_2$, P/2g\\
$\phi$  &  P/2g \\
\noalign{\smallskip}\hline
\end{tabular}
\end{center}
\end{table}

This model was extended to the case of electroproduction~\cite{Lag04}. 
%The longitudinal part of the leptonic and
%hadronic currents is related to the previously determined transverse
%components through gauge invariance.
The $Q^2$ dependence of the $f_2$ and P exchange is built in the model.
In the case of $\omega$ production, the only additional free parameters 
come from the electromagnetic form factor which accounts for the finite 
size of the vertex between the virtual photon, the exchanged $\pi^0$ trajectory
and the $\omega$ meson. 
This form factor could be chosen 
as the usual parameterization of the pion electromagnetic form factor: 
$F_{\omega\pi\gamma} = F_{\pi} = (1+Q^2/\Lambda_{\pi}^2)^{-1}$,
with $\Lambda_{\pi}^2= 0.462$ GeV$^2$. 
As described so far, the model fails to account for the observed
$t$ dependence (see dashed lines in fig.~\ref{fig:dsigmadtWfixe}).
From the observation that the differential cross section becomes nearly
$Q^2$-independent at high $-t$, an adhoc modification of the form factor
\begin{equation}
\label{eq:Fpiomegg2}
	F_{\omega\pi\gamma}(Q^2) \to  F_{\omega\pi\gamma}(Q^2,t) =
	\frac{1}{1+\frac{Q^2}{\Lambda_{\pi}^2}
		   \left(\frac{1+\alpha_{\pi}(t)}{1+\alpha_{\pi}(0)}\right)^2}
\end{equation}
was proposed~\cite{Lag04}.
The saturating $\pi^0$ Regge trajectory  obeys the
relation $\lim_{t\to -\infty}\alpha_{\pi}(t)= -1$,
so that the form factor becomes flat at high $-t$.
Thus, eq.~(\ref{eq:Fpiomegg2}) associates 
the point-like coupling of the virtual photon 
with the saturation of the $\pi^0$ Regge trajectory
which accounts for hard scattering in this kinematical limit~\cite{Lag04}.
Note that this modification of the form factor does not violate
gauge invariance, which holds separately for each contribution from 
Table~\ref{tab:JML} and, in the case of $\pi^0$ exchange, from the spin and 
momentum structure of the $\omega\pi\gamma$ vertex.

The $t$ dependence of the differential cross sections is then well described
(solid lines in fig.~\ref{fig:dsigmadtWfixe}). The $Q^2$ dependence
of the cross sections is illustrated in fig.~\ref{fig:xsec_Q2xB}. At
high $x_B$, which corresponds to the lowest values of $W$,
$s$-channel resonance contributions are not taken into account in the model
and may explain the observed disagreement. Finally the interference
terms $\sigma_{TT}$ and $\sigma_{TL}$ 
% are in qualitative agreement
agree in sign and trend, but not in magnitude,
with our results (fig.~\ref{fig:sigma_tt_tl}).

So, within this model, $\pi^0$ exchange, or rather the exchange of the
associated saturating Regge trajectory, continues to dominate the
cross section at high $Q^2$ and the cross section is mostly
transverse. This is consistent with our observations
of the dominance of unnatural parity exchange in the $t$-channel
in the previous section.
		  
%%%%%%%%%%%%%%%%%%%%%%%%%%%%%%%%%%%%%%%%%%%%%%%%%%%%%%%%%%%%%%%%%%%%
\section{Relevance of the handbag diagram}
\label{sec:handbag}
Let us recall that the handbag diagram of fig.~\ref{fig:handbag} is 
expected to be the leading one in the Bjorken regime. In this
picture, the transition $\gamma^*_L\to\omega_L$ would dominate the
process. This is clearly antinomic to the findings in
sect.~\ref{sec:regge}, where our results are interpreted as dominated
by the $\pi^0$ exchange, which is mostly due to transverse photons.
In addition, $\pi^0$ exchange is of a pseudo-scalar nature, while
the $H$ and $E$ GPD which enter the handbag diagram amplitude 
are of a vector nature.

Independent of the model interpretation presented in sect.~\ref{sec:regge}, 
our results point to the
non-conservation of helicity in the $s$-channel 
(figs.~\ref{fig:sigma_tt_tl}, \ref{fig:fitsVARphiN} and \ref{fig:rijMOMmeth3}),
meaning that the handbag diagram does not dominate the process, even for 
small values of $-t$ and $Q^2$ as large as 4.5 GeV$^2$.

\begin{figure}
\begin{center}
\resizebox{0.45\textwidth}{!}
	{ \includegraphics{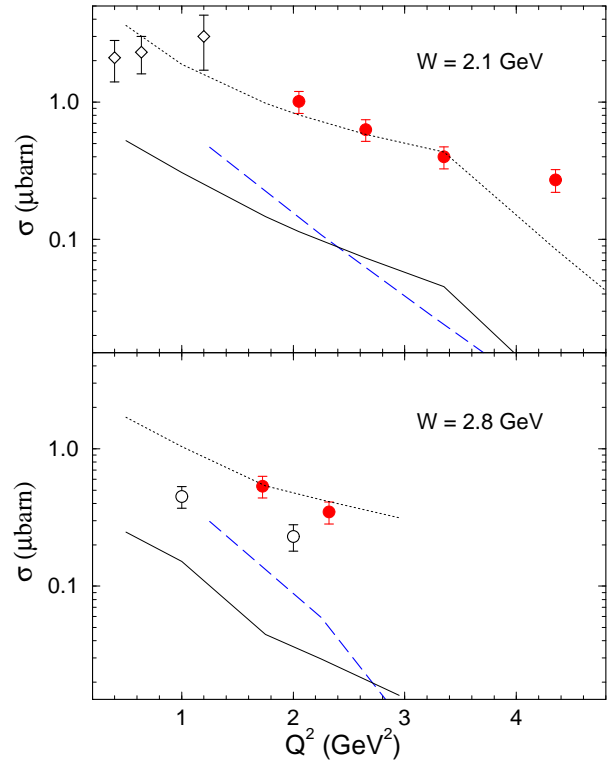}
	}
\caption{(Color online) 
	 Total cross sections 
	 for the reaction $\gamma^* p \to \omega p$,
	 for $\langle W\rangle =$ 2.1 (top) and 2.8 (bottom) GeV~: 
	 this work (full red circles), 
	 DESY data (empty diamonds), 
	 Cornell data (empty circles),
	 and JML model (dotted curves). 
	 The longitudinal contribution, $\varepsilon(E,Q^2)\sigma_L$,
	 is calculated  
	 according to the JML (solid lines) and VGG (dashed blue lines) models.
%	{\it Will be modified. Will add $\sigma$ curve of JML.}
	} 
\label{fig:VGG}
\end{center}
\end{figure}

As a consequence, $\sigma_L$ could not be extracted from our data
for a direct comparison with models based on the GPD formalism.
It is however instructive to consider here the predictions of a 
GPD-based model~\cite{Vdh97,Goe01}, denoted hereafter VGG.
This is a twist-2, leading order calculation, where the GPD are
parameterized in terms of double distributions ($DD$) and include the
so-called D-term (see Ref.~\cite{Goe01} for definitions):
$H,E \sim DD(x,\xi)e^{b(\xi,Q^2)t/2}$, where $b$ is taken from
the data (see sect.~\ref{subsec:xsec_t} and table~\ref{tab:sigma}).
An effective way of incorporating some of the higher twist effects
is to introduce a ``frozen" strong coupling constant $\alpha_S=0.56$.
This model is described in some more details in Ref.~\cite{Had04} and
is applied here to the specific case of $\omega$ production.
The model calculations (VGG and JML) of $\varepsilon\sigma_L$
are plotted in fig.~\ref{fig:VGG}. The sharp drop of the
curves at high $Q^2$ is due to the decrease of $\varepsilon$, at
our given beam energy, as $Q^2$ reaches its kinematical limit.
When compared to our results, $\varepsilon\sigma_L$ is 
calculated to be only
1/6 to 1/4 of the measured cross sections, thus explaining the difficulty 
in  extracting this contribution.

The $\omega$ channel thus appears to be a
challenging reaction channel to study the applicability
of the GPD formalism. This is attributed to the  $t$-channel $\pi^0$ 
exchange, which remains significant even at high values of $Q^2$.
In contradistinction, the $\pi^0$ exchange is negligible in the
case of the $\rho$ production channel, where SCHC was found
to hold, and $\sigma_L$ could be extracted and compared successfully
to GPD models~\cite{Air00,Had04}.

%%%%%%%%%%%%%%%%%%%%%%%%%%%%%%%%%%%%%%%%%%%%%%%%%%%%%%%%%%%%%%%%%%%%
\section{Summary}
\label{sec:sum}

An extensive set of data on exclusive $\omega$ electroproduction
has been presented, for $Q^2$ from 1.6 to 5.1 GeV$^2$ and
$W$ from 1.8 to 2.8 GeV ($x_B$ from 0.16 to 0.64). Total and
differential cross sections for the reaction $\gamma^*p\to\omega p$
were extracted, as well as matrix elements linked to the
$\omega$ spin density matrix.

The $t$ differential cross sections are surprisingly large for high
values of $-t$ (up to 2.7 GeV$^2$). This feature can be accounted
for in a Regge-based model (JML), provided a $t$ dependence 
is assumed for the $\omega\pi\gamma$ vertex form factor, with a
prescription inspired from saturating Regge trajectories. 
%The full meaning of this parameterization should be investigated
%in more detail. 
It appears that the virtual photon is more likely 
to couple to a point-like object as $-t$ increases.

The analysis of the $\phi$ differential cross sections and of the
$\omega$ decay matrix elements indicate that the $s$-channel helicity
is not conserved in this process. As a first consequence, the longitudinal
and transverse contributions to the cross sections could not be
separated. Furthermore, the values of some decay matrix elements
point to the importance of unnatural parity exchange in the
$t$-channel, such as $\pi^0$ exchange. 
This behaviour had been previously established in the case
of $\omega$ photoproduction, but
not for the large photon virtuality obtained 
in this experiment.
The results on these observables also support the
JML model, where the exchange of the saturating Regge trajectory
associated with the $\pi^0$ is mostly transverse and dominates the process.

Finally,
the experiment demonstrated that exclusive vector meson electroproduction
can be measured with high statistics in a wide kinematical range.
The limitations at high $Q^2$ were not due to the available luminosity
of the CEBAF accelerator or to the characteristics of the CLAS spectrometer, 
but to the present beam energy. With
the planned upgrade of the beam energy up to 12 GeV~\cite{upgrade},
such reactions will be measured to still higher values of $Q^2$.
In the specific case of the $\omega$ meson, 
as was shown in this paper, this will be a necessary condition
for the extraction of a longitudinal contribution
of the handbag type, 
related at low values of $-t$ to generalized parton distributions.
More generally, this experiment opens a window on the high $Q^2$
and high $-t$ behaviour of exclusive reactions, which needs
further exploration.
%%%%%%%%%%%%%%%%%%%%%%%%%%%%%%%%%%%%%%%%%%%%%%%%%%%%%%%%%%%%%%%%%%%%
\begin{acknowledgement}
We would like to acknowledge the outstanding efforts of the staff of the 
Accelerator and the Physics Divisions at JLab that made this experiment possible.
This work was supported in part by 
the Italian Istituto Nazionale di Fisica Nucleare, 
the French Centre National de la Recherche Scientifique, 
the French Commissariat \`{a} l'Energie Atomique, 
the U.S. Department of Energy and National Science Foundation, 
the Emmy Noether grant from the Deutsche Forschungs Gemeinschaft 
and the Korean Science and Engineering Foundation.
The Southeastern Universities Research Association (SURA) operates the 
Thomas Jefferson National Accelerator Facility for the United States 
Department of Energy under contract DE-AC05-84ER\-40150.
\end{acknowledgement} 
%%%%%%%%%%%%%%%%%%%%%%%%%%%%%%%%%%%%%%%%%%%%%%%%%%%%%%%%%%%%%%%%%%%%

\end{document}